\pdfoutput=1 %
\documentclass[3p,review,authoryear]{elsarticle} 
\usepackage{natbib}
\usepackage{amsmath} 
\usepackage{siunitx}
\usepackage{rotating}
\usepackage{booktabs}
\usepackage{xcolor}

\biboptions{longnamesfirst,semicolon}
\title{Climate of Earth-like planets with high obliquity and eccentric orbits: implications for habitability conditions}

\author[mi]{Manuel Linsenmeier\corref{cor1}} \ead[url]{manuel.linsenmeier@zmaw.de} \author[mi]{Salvatore Pascale} \author[mi,uk,uk2]{Valerio Lucarini} \cortext[cor1]{Corresponding author\\ {\it E-mail address}: manuel.linsenmeier@zmaw.de} \address[mi]{KlimaCampus, Meteorologisches Institut, Universit\"at Hamburg, Hamburg, Germany} \address[uk]{Department of Mathematics and Statistics, University of Reading, Reading, UK} \address[uk2]{Walker Institute for Climate System Research, University of Reading, Reading, UK}

\begin{document} 

\begin{abstract} We explore the effects of seasonal variability for the climate of Earth-like planets as determined by the two parameters polar obliquity and orbital eccentricity using a general circulation model of intermediate complexity. In the first part of the paper we examine the consequences of different values of obliquity and eccentricity for the spatio-temporal patterns of radiation and surface temperatures as well as for the main characteristics of the atmospheric circulations. In the second part we analyse the associated implications for the habitability of planets close to the outer edge of the habitable zone (HZ). This part of the paper focuses in particular on the multistability property of climate, i.e. the parallel existence of both an ice-free and an ice-covered climate state. Our results show that seasonal variability affects both the existence of and transitions between the two climate states. Moreover, our experiments reveal that planets with Earth-like atmospheres and high seasonal variability can have ice-free areas at much larger distance from the host star than planets without seasonal variability, which leads to a substantial expansion of the outer edge of the HZ. Sensitivity experiments exploring the role of azimuthal obliquity and surface heat capacity test the robustness of our results. On circular orbits, our findings obtained with a general circulation model agree well with previous studies based on one dimensional energy balance models, whereas significant differences are found on eccentric orbits. \end{abstract} 


\maketitle

\section{Introduction}\label{sec:intro} 

Over the last two decades, more than 1000 planets outside our Solar System (exoplanets) have been confirmed and several thousands are classified as candidates \citep{Udry_2007,Borucki_2011}. Beyond the sheer detection, observation techniques based on the radial velocity and transit method also allow one to infer some planetary properties such as mass, radius, and orbit. Today, much research is devoted to the interpretation of the statistical distribution of these properties \citep{Howard_2013}.

One of the current research questions in the field of exoplanets is how different orbital and planetary parameters affect the habitability of a planet, i.e. its ability to host life \citep{Seager_2013}. Commonly, a planet is considered habitable if its surface conditions allow for the existence of liquid water. The estimation of habitability is then usually provided as an estimation of the habitable zone (HZ), i.e. the range of distances from the host star that would allow for habitable conditions. Consequently, two different processes define the inner and outer boundaries of the HZ: at the inner boundary, the run-away greenhouse effect leads to a complete evaporation of water at the surface; at the outer boundary, a completely frozen surface limits the habitability of a planet \citep{Hart_1979,Kasting_1993}.

Climate models can provide insights  to the investigation of exoplanets and their habitability. Ranging from toy models to general circulation models, they allow one to study the effect of different processes with variable degree of approximation and computational cost. Radiative-convection models (RCMs) were the first climate models applied to the investigation of exoplanets (e.g. \citet{Kasting_1993}). In these one dimensional vertical models, the atmosphere of the planet is treated as a single column. As they do not capture effects of temporal or spatial heterogeneity, they were subsequently complemeted by latitudinally resolved energy-balance models (EBMs) (e.g. \citet{Williams_1997}). More recently, also general circulation models (GCMs) have been employed in habitability studies and, more generally, to study their climates and circulations \citep{Merlis_2010,Pierrehumbert_2011a,Heng_2011,Heng,Menou_2012,Showman_2013}.

Estimates of the HZ differ among these models, as they include different processes at different degree of approximation. Simulations with RCMs yield an extent of the HZ of an Earth-like planet orbiting around a Sun-like star from \SI{0.99}{AU} to \SI{1.70}{AU} \citep{Kopparapu_2013}. Results obtained from EBMs and GCMs, however, indicate that RCMs tend to underestimate the extent of the HZ due to processes that are not or not sufficiently represented in these models. Simulations with EBMs show that the inclusion of seasonal variability due to either a non-zero planetary obliquity, an eccentric orbit or both can extend the outer boundary of the HZ relative to results from RCMs \citep{Williams_1997,Spiegel_2009,Dressing_2010}. As recent work shows, the extension of the HZ can be further increased if planetary obliquity and orbital eccentricites vary over time \citep{Armstrong_2014}. Moreover, results from a GCM show that a better representation of cloud feedbacks can lead to an extension of the HZ relative to results from RCMs also at its inner boundary \citep{Leconte_2013,Yang_2013}.

\begin{figure}[htb!]
\centering \includegraphics[width=0.49\textwidth]{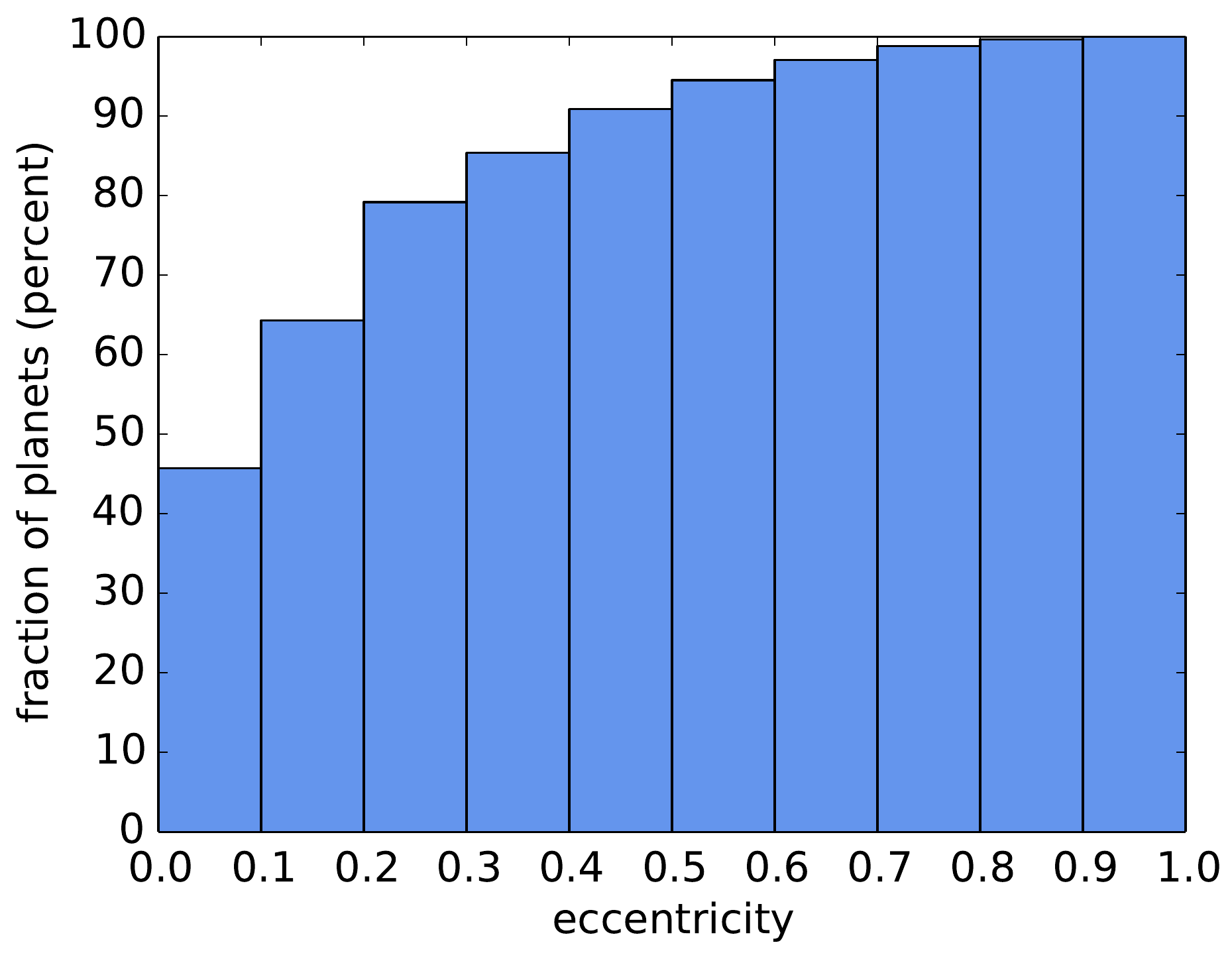}
\caption{Cumulative histogram of orbit eccentricities of 855 exoplanets whose eccentricity were estimated; the data were retrieved from the Extrasolar Planets Encyclopedia on October 5th 2014 \citep{Exoplanet_www}}
\label{fig:ecc}
\end{figure}

Large seasonal variability might indeed be a common feature of extra-solar planets as detected planets exhibit a large diversity of orbital eccentricities \citep{Ford_2008,Moorhead_2011,Kane_2012b}. Around 35 \% of the detected exoplanets are located on an orbit with an eccentricity $e > 0.2$ and around 9\,\% on an orbit with $e > 0.5$ (Figure \ref{fig:ecc}). Highly eccentric orbits feature dramatic intra-annual variability: while for $e=0.2$ a planet at periastron receives about twice the amount of energy at apoastron, this factor increases to around 9 for $e = 0.5$ \citep{Dressing_2010}. Also the polar obliquity of a planet $\theta$ plays an important role in determining seasonal variability \citep{Spiegel_2009,Dobrovolskis_2013}. In particular the combination of a high value of $e$ and a high value of $\theta$ can lead to extreme seasonal variability \citep{Dressing_2010}.

Multistability of the climate state further complicates estimates of the outer boundary of the HZ. Since the experience gathered with the first EBMs, it is well known that the climate of a planet can exhibit two stable states with different degree of ice-coverage \citep{Budyko_1969,Sellers_1969}, usually referred to as warm (ice-free or only partially ice-covered) and cold (completely or mostly ice-covered) climate state. Evidence suggests that Earth might have entered and exited a cold state several times in the past \citep{Hoffman_2002}, and the transition to the cold state can be reproduced with state-of-the-art GCMs \citep{Marotzke_2007,Voigt_2010,Voigt_2011}. While the multistability property has been extensively discussed in the context of paleoclimate (see also \citet{Romanova_2006,Yang_2012,Yang_2012b}), its implications for the habitability of exoplanets is a relatively new area of research. \citet{Lucarini_2010} provided an investigation of bistability in a thermodynamic framework, which has recently been extended to include the parameters rotation rate and atmospheric opacity \citep{Boschi_2013,Lucarini_2013}.

The effect of obliquity and eccentricity has so far primarily been investigated with EBMs \citep{Williams_1997,Spiegel_2009,Dressing_2010}. Their results reveal that both parameters have substantial implications for the extent of the HZ, particularly on eccentric orbits \citep{Spiegel_2009,Dressing_2010}. As downside of the numerical efficiency of these models, however, they rely on crude simplifications. One of their most critical simplifications is that ice is represented by simply assuming a temperature dependent surface albedo. Moreover, they do not include any dynamical processes apart from a simple diffusion parametrization of meridional  heat transport. However  \cite{Ferreira_2014} show that the phenomenological  transport efficiency relating temperature gradients and heat transport varies by an order of magnitude amongst earth-like planets at different obliquities.  As these  parameterizations used in EBMs for meridional heat transport are tuned to reproduce the climate of Earth, their validity is challenged by the very different shapes of atmospheric circulations and climates that can be expected on exoplanets with high polar obliquities, highly eccentric orbits, and variable degree of ice-coverage.

Despite the need for simulations with climate models of higher complexity, only three studies assessed the effect of either obliquity or  eccentricity on atmospheric circulations and climate with a GCM so far \citep{Williams_2002,Williams_2003,Ferreira_2014}. They did, however, not provide a systematic investigation of the implications for habitability. This is the aim of this work. Using a GCM of intermediate complexity we explore the atmospheric circulations and climates of idealized Earth-like planets for different astronomical parameters and examine the implications for habitability.

It is well known that silicate weathering affects the climate of Earth as a stabilizing mechanism acting on timescales of million of years: when surface temperatures are sufficiently low and rocks are not exposed to the atmosphere, the weathering is strongly suppressed, with an ensuing accumulation of CO2 in the atmosphere \citep{Walker_1981}. In our study, we do not consider this process (see instead \citet{Williams_1997}) but keep the CO2 concentration at a fixed value, following \citet{Spiegel_2009} and \citet{Dressing_2010}. Therefore, our study should be thought primarily as a parametric exploration of the effect of changing orbital parameters on a planet of given atmospheric composition rather than a realistic study of the HZ.

We anticipate that seasonal variability leads to an increase of the the maximal distance between planet and host star that allows for habitable conditions. The combined effect of obliquity and eccentricity on multistability has to our knowledge not been investigated yet. Our results reveal that obliquity is crucial in determining the extent of both the warm and cold state. Moreover, eccentric orbits are generally associated with a narrower range of two  stable climate states. Our simulations also show that seasonal variability primarily leads to temporally ice-free regions, which stresses the importance of different definitions of habitability. Sensitivity experiments that include the parameters azimuthal obliquity and ocean heat capacity test the robustness of our results. Since we neglect the effect of silicate weathering, our results of the outer boundary of the habitable zone can only be used as conservative estimates. Large intra-annual variations of temperature and ice-coverage found in many of our experiments however question traditional estimates with energy balance models that do not take these variations into account.

The paper is structured as follows. In Section \ref{sec:model} we introduce and discuss our model PlaSim, explain the experimental setup, and list all simulations. The main features of atmospheric circulations of planets at different obliquities and eccentricities are presented in Section \ref{sec:climates}. Section \ref{sec:results} shows the implications of obliquity and eccentricity for multistability and the degree of habitability. A discussion of our results follows in Section \ref{sec:discuss}. Finally, we draw our conclusions in Section \ref{sec:concl}.

\section{Model and experimental setup}\label{sec:model}

\subsection{The Planet Simulator} 

All numerical simulations are performed with the Planet Simulator (PlaSim), a general circulation model of intermediate complexity \citep{Fraedrich_2005},  freely available at {\emph http://www.mi.uni-hamburg.de/plasim}. Atmospheric processes included in PlaSim are: atmospheric dynamics, surface turbulent fluxes, clear-sky and cloudy-sky atmosphere-radiation interaction, and moist and dry convection. The atmospheric component  of the model is run coupled to a thermodynamic sea ice model and a slab ocean model (50m). The model excludes any sea ice and ocean dynamics as well as tracer chemistry and vegetation dynamics. In the following we briefly describe the parameterizations of the atmospheric component of the model that are included in the current set of numerical simulations.

\begin{itemize}

\item  \emph{Dynamic Solver.} PlaSim relies on a  spectral atmospheric dynamical core  to integrate the primitive equations for the vertical component  of the vorticity $ \zeta=\mathbf k\cdot(\nabla\times\mathbf{v})$ and horizontal wind divergence $\delta=\nabla_h\cdot\mathbf{v}$, temperature $T$ and surface pressure $p_s$.  In spectral dynamical solvers the prognostic variables are represented as a sum of spherical harmonics truncated at order $N$ (triangular truncation is implemented in our case). Due to truncation, numerical hyperviscosity   $\sim  \nabla^8 (\zeta, \delta, T)$ is  applied in PlaSim to $\zeta, \delta, T$ to remove the enstrophy accumulating at the smallest resolved scales  generating numerical instability.

Sound waves are filtered by hydrostatic approximation. The Robert-Asselin time filter is used to suppress spurious computational modes associated with the leapfrog time-stepping scheme implemented to deal with the vorticity equation. More details about the numerical methods can be found in \cite{Lunkeit_2011}.

\item \emph{Radiation.} The shortwave radiation scheme is based on  \cite{Lacis_1974} for the clear-sky atmosphere. The solar spectral range is divided into a visible and ultraviolet part  $\lambda < 0.75$ $\mu$\,m with  ozone absorption and Rayleigh scattering and a near infrared part  $\lambda > 0.75$ $\mu$\,m  with water vapor absorption only.  For the cloudy fraction, transmissivities and albedos, both depending on the cloud liquid water content, are calculated differently in the near infrared and visible-ultraviolet bands, following the ideas of  \cite{Stephens_1978} and \cite{Stephens_1984}. The incoming solar flux $F_{sw}^{toa}$ at the top of the atmosphere is $F_{sw}^{toa}=S^\star \cos Z$, where $S^\star$ is the solar constant and $Z$ the zenith angle, a function of the latitude and time estimated according to  \cite{Berger_1978}. 

Longwave radiation for the clear sky relies on a broad band emissivity method \citep[e.g. ][]{Rodgers}.  The transmissivities for water vapor, carbon dioxide  and ozone  depend on their local concentration and are determined according to the empirical relations of \cite{Sasamori_1968}.  For each  layer with a certain fractional cloud cover, the longwave transmissivity and cloud albedo  are determined from  the cloud liquid water content, the clear-sky transmissivity and the fractional cloud cover,  as discussed  in  \citet{Slingo_1991}.   For more extensive details of the equations used to parametrize the longwave and shortwave transmissivities the reader is referred to \cite{Lunkeit_2011}.

From the optical properties of the atmosphere (shortwave and longwave transmissivity and albedo), $F_{sw}^{toa}$ and the surface temperature upwards and downwards radiative fluxes at the interfaces of each atmospheric layer and hence the radiative heating rates are computed.

\item \emph{Surface and moist processes.} In all simulations the lower boundary is a flat surface with prescribed albedo and heat capacity. This is implemented with a shallow slab ocean model. The surface temperature therefore  evolves in time according to  $C_{slab}\dot{T}_s=F_{sw}^{surf}+F_{lw}^{-}=\sigma_B T_s^4+F_{sens}+F_{lat}$, where $C_{slab}$ is the  heat capacity of the slab ocean,  $F_{sw}^{surf}$ is the net solar radiation at the surface, $F_{lw}^{-}$ the downward longwave radiation at the surface and $F_{sens}+F_{lat}$ the sum of sensible and latent heat fluxes at the surface.

Cumulus convection is parameterized by a Kuo-type convection scheme  \citep{Kuo_1965,Kuo_1974}. Large scale condensation occurs when the air is supersaturated. Because condensed water is instantaneously removed as precipitation, no water is stored in clouds. Cloud cover and cloud liquid water content are diagnostic quantities. The fractional cloud cover of a grid box is parameterized following \cite{Slingo_1991} using the relative humidity for the stratiform cloud amount and the convective precipitation rate for the convective cloud amount.  The bulk aerodynamic formulas are employed for surface flux parameterizations of wind stress ($\tau_x, \tau_y$), sensible heat flux  $F_{sens}$ and latent heat flux $F_{lat}$. Drag and transfer coefficients follow the approach by \cite{Louis_1979} and \cite{Louis_1981}.

\end{itemize}

The thermodynamic sea ice model is based on the 0-layer version of the model of \citet{Semtner_1976}, which is a simplified version of the \citet{Maykut_1971} thermodynamic sea ice model. Its main simplifications are the exclusion of the capacity of the ice to store heat and the assumption of a constant temperature gradient in the ice sheet (note that for the computation of the sea ice surface temperature a heat capacity corresponding to an ice layer with a thickness of 10 cm is assumed). Despite these simplifications, a comparison with both a more sophisticated 3-layer version of the same model and the Maykut and Untersteiner model shows deviations in the seasonal cycle of ice thickness of only about tens of centimeters \citep{Semtner_1976,Semtner_1984}. Further simplifications of the model are the exclusion of oceanic heat transport and the exclusion of wind induced sea ice drift.

The thickness of sea ice in equilibrium is governed by thermodynamic processes (freezing and melting) and dynamic processes (advection, collision and deformation). Only thermodynamic processes are included in our model. The ice thickness is then constrained by the diffusion of heat through the ice sheet, which primarily depends on the surface temperature and oceanic heat flux. In order to avoid artificial sources of energy, the oceanic heat flux is set to zero in our model. Its limiting effect on ice growth is accounted for by prescribing a maximal ice thickness $h_{i, \rm{max}} = \SI{3}{m}$. This value is in good agreement with observation of equilibrium ice thickness between 2 and 3 m of 1-year and 2-year old sea ice on Earth \citep{Haas_2009}. A detailed description of the sea ice model can be found in the PlaSim reference manual \citep{Lunkeit_2011}. A more detailed discussion of the model and also on neglecting dynamic sea ice processes can be found in \citet{Ebert_1993}.

Depending on the ice model, different processes are involved in the ice-albedo feedback. The simplest form of the feedback captures only changes in the areal extent of the ice cover. An initial perturbation of surface temperature is then amplified or damped only if it leads to a change of the the area that is ice-covered and ice-free. More realistically, also further processes associated with the ice thickness, snow cover, or melt pond characteristics can constitute positive or negative feedback mechanisms \citep{Curry_1995}. Yet these processes require more sophisticated models and are therefore not taken into account in our experiments.

PlaSim offers a reasonable trade-off between numerical efficiency and the number of explicitly included processes. Even though all model elements have been tuned to simulate the Earth's climate, we expect them to remain reasonably accurate for climates close to Earth's  climatic conditions. In addition to have been intensively used to study the climate of Earth, the model has already been employed in investigations of atmospheric circulations of terrestrial planets \citep{Stenzel_2007,Pascale_2013}, snowball transitions for different astronomical and atmospheric characteristics \citep{Lucarini_2013, Boschi_2013} and the climate of tidally-locked, water-trapped planets \citep{Menou_2013}. 

\subsection{Astronomical parameters: obliquity and eccentricity}

The two astronomical parameters investigated in this work are the obliquity $\theta$ of a planet (the tilt of its spin axis with respect to a normal to the plane of the orbit) ranging from $\SI{0}{\degree}$ to $\SI{90}{\degree}$ and the eccentricity $e$ of its orbit ranging from 0 to 1. Both parameters determine the temporal and spatial distribution of solar insolation (see \citet{Berger_1978} for a comprehensive derivation of solar insolation as function of these two parameters; spatio-temporal patterns of insolation are also explored in \citet{Dobrovolskis_2013}).

\begin{figure}[htb!]
\centering \includegraphics[width=0.49\textwidth]{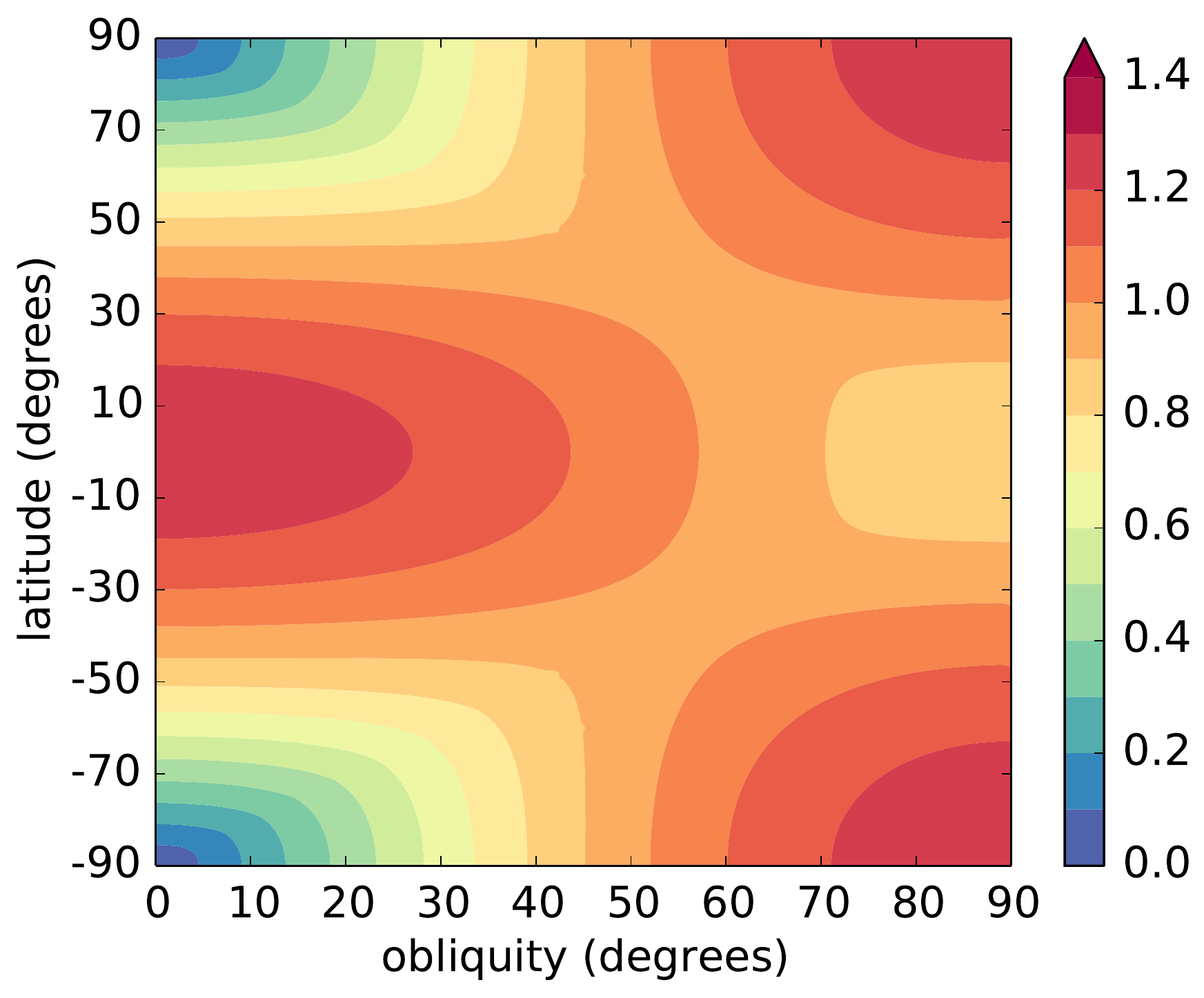}
\caption{Annual and zonal mean irradiation at the top of the atmosphere depending on latitude and the obliquity of the planet; shown as fraction of the global average insolation $S/4$ on a circular orbit (eccentricity $e=0$)}
\label{fig:ins}
\end{figure}

On circular orbits the spatial distribution of annual mean irradiation at the top of the atmosphere is determined by the obliquity of the planet (Figure \ref{fig:ins}). The relation between irradiation at low and high latitudes is reversed at the critical value $\theta_{c} \approx \SI{55}{\degree}$. For $\theta \leq \theta_{c}$, low latitudes receive on annual average more radiation than high latitudes. The converse applies for $\theta > \theta_{c}$ (Figure \ref{fig:ins}). This consequence of high obliquity has also been considered relevant for explaining geological records that hint at low-latitude glaciation in Earth's history (\citet{Williams_1998,Williams_2003}; see also \citet{Hoffman_2002} for an overview), although no physical mechanism responsible for large changes of Earth's obliquity has been found \citep{Levrard_2003, Pierrehumbert_2011b}.

In our experiments we set $\theta$ to $\SI{0}{\degree},\SI{60}{\degree}$, and $\SI{90}{\degree}$, representing planets with most intensive irradiation at the equator, weak latitudinal gradients, and the highest annual mean insolation at the poles, respectively. As we aim at a parametric investigation of how different values of obliquity influence climate, we do not account for the effect of tidal processes that can erode the obliquity of a planet on time scales of billions of years \citep{Heller_2011}.

On eccentric orbits, the insolation pattern is also influenced by the orbital longitude of periastron $\omega$ measured from northern vernal equinox (also refered to as azimuthal obliquity). This third astronomical parameter determines the hemispheric asymmetry of the annual mean insolation. Most of our analysis deals with planets with $\omega = \SI{0}{\degree}$, which feature an annual mean irradiation symmetric with respect to the equator. In other words, periastron and apoastron are aligned with vernal equinoxes and both hemispheres receive the same amount of energy per year. On Earth, $\omega$ changes over time, with a current value of $\omega \approx \SI{103}{\degree}$ \citep{Berger_1978}. In order to test the sensitivity of our results against changes in $\omega$, we conduct two simulations with  $\omega = \SI{30}{\degree}$ and $\omega = \SI{90}{\degree}$. The latter is equivalent to an alignment of periastron with solstice, thus maximizing the difference of mean irradiation between the northern and southern hemisphere.

\subsection{Experimental setup}\label{sec:model:planet}

All simulations are performed with a horizontal resolution of 64 $\times$ 32 grid points (T21 spectral resolution) and 10 vertical levels, which enables us to run a large ensemble of numerical simulations in a reasonable amount of time. In experiments not included in this paper the T21 horizontal resolution - corresponding approximately to $500$ km - has proven to provide an accurate representation of the large scale circulation features as compared to higher resolutions (T42).  Most importantly, the T21 horizontal resolution captures baroclinic eddies, which play a fundamental role in the transport of energy and momentum polewards \footnote{The Rossby deformation radius $L_R= N H_{t}/f$ \citep{Eady, James} is a good approximation for the size of these eddies. For a rotation rate equal to Earth's one,  $L_R \sim 1000$ km, with the Coriolis parameter at mid-latitudes $f\approx 10^{-4}$ rad\,s$^{-1}$, the buoyancy frequency $N\sim 10^{-2}$  s$^{-1}$ and the height scale $H_t\sim 10$ km.}. Direct comparison between runs obtained at the two horizontal resolution T21 and T42 (not shown) demonstrate that all main features (jet streams, meridional circulation, temperature structure) are well captured at T21.  Similar experiments showed that the vertical resolution ensures a convergence of the radiative transfer and an adequate representation of the vertical structure of the atmosphere. 

Although the model PlaSim has been used to study the climate of planets with very different properties (see for instance \citet{Stenzel_2007} for an investigation of the climate of Mars), in this work we focus on idealized planets that are Earth-like with respect to their radius, density, host star, and atmospheric composition. This choice of the setup is motivated by the presence of several studies on snowball dynamics and habitability for Earth-like planets, employing both simple EBMs \citep{Budyko_1969,Sellers_1969,Ghil_1976,Bodai_2014} and GCMs \citep{Romanova_2006,Marotzke_2007,Voigt_2010,Voigt_2011,Lucarini_2010}, which allows us to build on a broad basis and put our analysis in a wider context.

All planets studied in this work are completely covered with water, i.e. aquaplanets. Therefore, surface properties such as surface roughness, moisture, and heat capacity are fixed and uniform all over the globe. Also the albedo of snow- and ice-free areas is uniformly set to $\alpha = 0.069$, as corresponding to an ice-free ocean. The consequences of different land-ocean distributions or different surface materials are therefore not considered in our experiments. We acknowledge that these properties might however play a crucial role in determining spatial and temporal variability and therefore interact with orbital forcing in a complex way. For instance, a planet with one large North Polar continent can experience large temperature oscillations for highly tilted spin axes due to lower heat capacity of land than ocean, potentially leading to the formation of ice caps. \citet{Spiegel_2009} show that the climate of such planet indeed also shows a strong dependence on obliquity and that its habitability exhibits a much more seasonal character than on a uniform planet. A list of  parameters and their chosen values can be found in Table \ref{tab:params}. 

\begin{table*}[tb]\centering\footnotesize
\begin{tabular}{ l  l  r  l} 
\toprule Parameter & Symbol in text & Value(s) & Units  \\ \midrule planet radius                       &                   &                    6300 &                    km \\ 
rotation rate                       &                   & 7.29 $\times$10$^{-5}$ &            rad s$^{-1}$ \\ 
gravitational acceleration          &                   &                    9.81 &            m s$^{-2}$ \\ polar obliquity                     & $\theta$          &           \{0, 60, 90\} &               degrees \\ orbit eccentricity                  & $e$               &              \{0, 0.5\} &                       \\ azimuthal obliquity                 & $\omega$          & $^{\star}$\{0, 30, 90\} &               degrees \\ CO$_{2}$ concentration              &                   &                     380 &                   ppm \\ length of one year                  & $T$               &                     360 &                     d \\ length of one day                   &                   &                      24 &                     h \\ ocean albedo                        & $\alpha$          &                  0.069 &                       \\ ocean depth                         & $d_{\rm{ocean}}$  &   $^{\star}$\{10, 50\} &                     m \\ ocean water density                 &                   &                    1030 &           kg m$^{-3}$ \\ ocean mass specific heat capacity   &                   &                    4180 &  J kg$^{-1}$ K$^{-1}$ \\  ocean horizontal diffusivity        &                   &                     0.0 &      m$^{2}$ s$^{-1}$ \\ sea ice albedo                      &                   &                     0.7 &                       \\ sea ice density                     &                   &                     920 &           kg m$^{-3}$ \\ sea ice mass specific heat capacity &                   &                    2070 &  J kg$^{-1}$ K$^{-1}$ \\ sea ice heat conductivity           &                   &                    2.03 &   W m$^{-1}$ K$^{-1}$ \\ sea ice minimal thickness           &                   &                     0.1 &                     m \\ sea ice maximal thickness           & $h_{i, \rm{max}}$ &      9 &                     m \\ \bottomrule \end{tabular} \caption{Parameters and their values of all experiments in this study; $^{\star}$denotes parameters that are considered in sensitivity experiments}
\label{tab:params} 
\end{table*}

We modulate the intensity of the stellar irradiation $S$ that reaches the top of the atmosphere in all our simulations, which can alternatively be interpreted as changing the planet-star distance. As our model is not able to simulate the physical properties of a thick, warm atmosphere with a substantial mass fraction of H$_{\rm{2}}$O, we limit our analysis to the outer edge of the HZ. Limitations of our model also imply that our definition of habitability differs from the maximum greenhouse limit that is usually considered as the outer boundary of the HZ \citep{Kasting_1993}. Our model does not account for the effects of silicate weathering, i.e. the stabilizing feedback of CO$_{\rm{2}}$ accumulation in the atmosphere, acting on timescales of millions of years, once the temperature falls sufficiently low to affect carbon fluxes at the surface. This implies that our study, in fact, should be thought primarily as a parametric exploration of the effect of changing orbital parameters on a planet of given atmospheric composition. Since we neglect a feedback pushing climate towards conditions favourable for life, we deduce that our estimated extents of the HZ are always conservative providing a lower boundary for the outer edge of the HZ. In the following, whenever we use the term habitable zone (HZ) we refer to this conservative definition of habitability. The limitations of our investigation, in particular regarding the lack of representation of the silicate weathering cycle, are later discussed in Section 5.

In the following we refer to the irradiation relative to the irradiation on Earth as $\tilde{S} = S/S^{\star}$ with $S^{\star} = \SI{1365}{Wm^{-2}}$. A list of all experiments whose results are presented in this work is shown in Table \ref{tab:sims}. 

\begin{table*}[tb]\centering \footnotesize
\begin{tabular}{ l  r  r  r  r  r } \toprule & \multicolumn{5}{l}{Parameter values} \\ \cmidrule{2-6} Experiment & $e$ & $\theta$ & $\omega$ & $d_{\rm{ocean}}$ & $h_{i,\rm{max}}$ \\ \midrule E00o00       & 0   & $ 0^{\circ}$ & $ 0^{\circ}$ &  50 m & 3 m \\ E00o60       & 0   & $60^{\circ}$ & $ 0^{\circ}$ &  50 m & 3 m \\ E00o90       & 0   & $90^{\circ}$ & $ 0^{\circ}$ &  50 m & 3 m \\ E02o00       & 0.2 & $ 0^{\circ}$ & $ 0^{\circ}$ &  50 m & 3 m \\ E02o60       & 0.2 & $60^{\circ}$ & $ 0^{\circ}$ &  50 m & 3 m \\ E02o90       & 0.2 & $90^{\circ}$ & $ 0^{\circ}$ &  50 m & 3 m \\ E05o00       & 0.5 & $ 0^{\circ}$ & $ 0^{\circ}$ &  50 m & 3 m \\ E05o60       & 0.5 & $60^{\circ}$ & $ 0^{\circ}$ &  50 m & 3 m \\ E05o90       & 0.5 & $90^{\circ}$ & $ 0^{\circ}$ &  50 m & 3 m \\ \midrule E05o90om30  & 0.5 & $90^{\circ}$ & $30^{\circ}$ &  50 m & 3 m \\ E05o90om90  & 0.5 & $90^{\circ}$ & $90^{\circ}$ &  50 m & 3 m \\ E05o90oc10  & 0.5 & $90^{\circ}$ & $ 0^{\circ}$ &  10 m & 3 m \\ \bottomrule \end{tabular} \caption{List of experiments and the respective values of the parameters eccentricity $e$, obliquity $\theta$, azimuthal obliquity $\omega$, ocean depth $d_{\rm{ocean}}$, and maximal sea ice thickness $h_{i,\rm{max}}$}
\label{tab:sims}
\end{table*}

\section{Climate states and atmospheric circulations} \label{sec:climates}

\begin{figure*}[htb!]
\centering \includegraphics[width=0.49\textwidth]{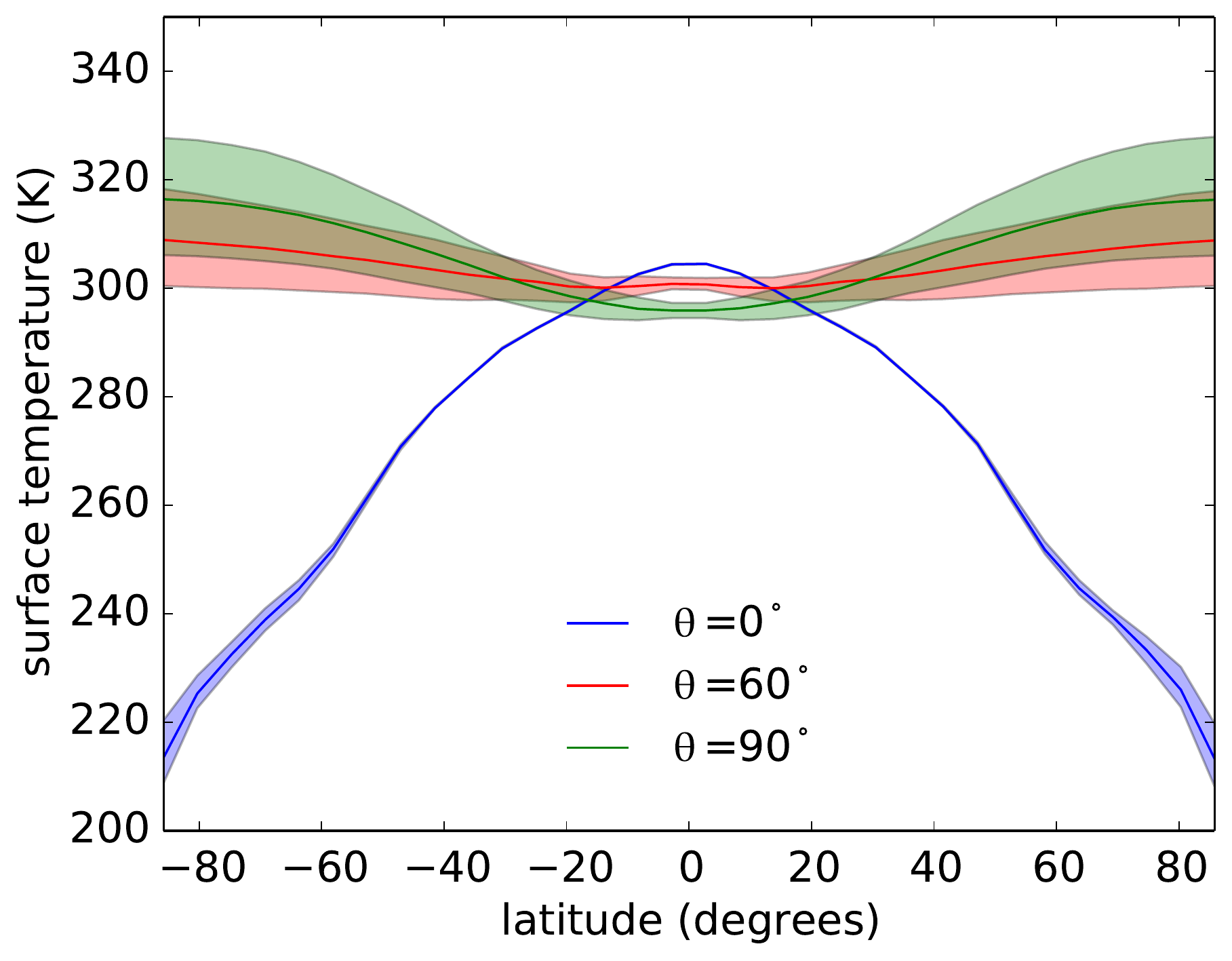}
\includegraphics[width=0.49\textwidth]{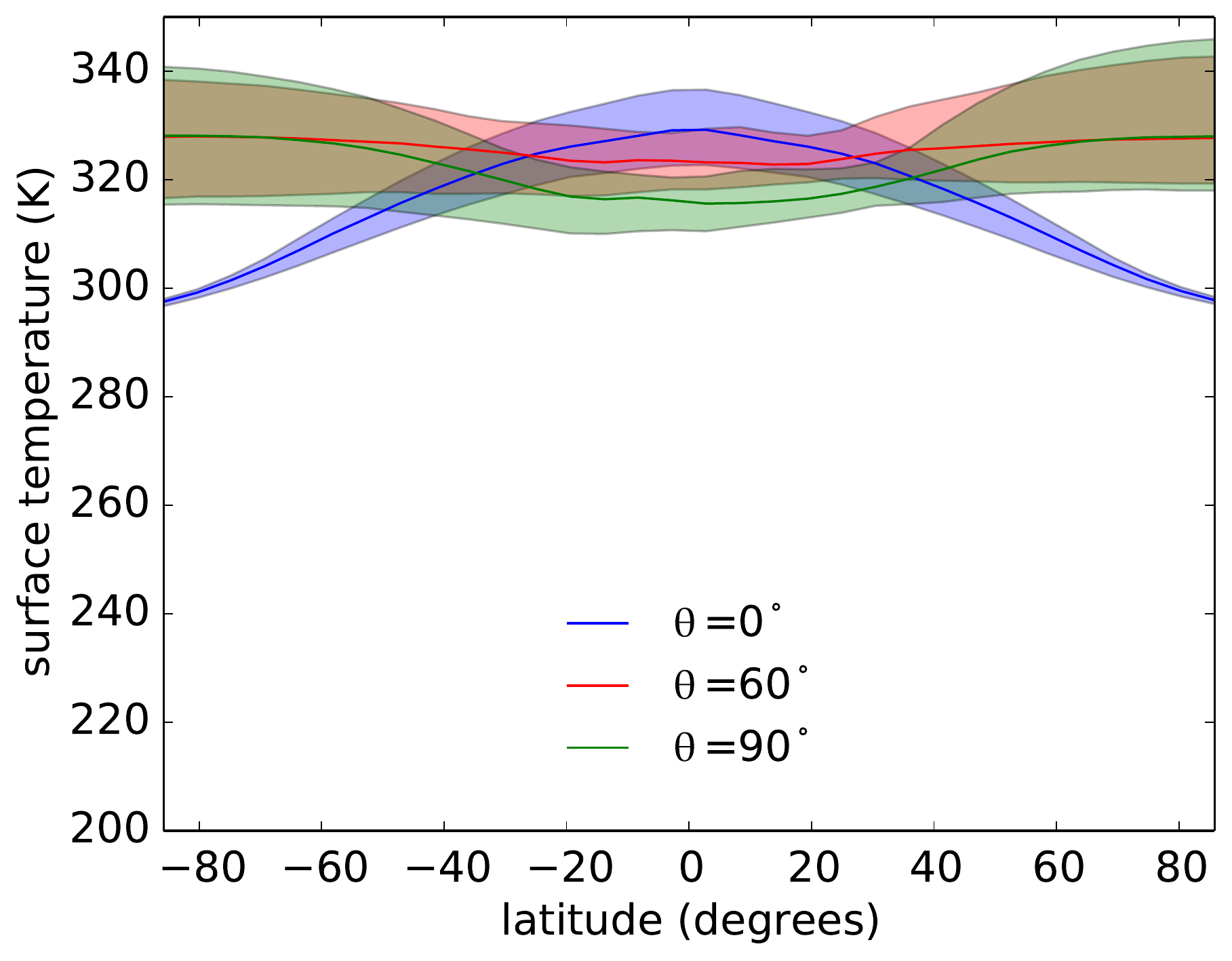}
\caption{Annual mean, minimum and maximum surface temperature of planets with different $\theta$ and an orbit with (left) $e=0$ and (right) $e=0.5$; irradiation $S = S^{\star}$ in all experiments; (left) only the warm state (upper branch of Figure \ref{fig:bifurc}) is shown}
\label{fig:gradient}
\end{figure*}

The annual and zonal mean surface temperatures of simulations with $\theta=0^\circ,\,60^\circ,\,90^\circ$ and $e=0,\,0.5$ for ice-free climates at $S=S^\star$ are shown in Figure \ref{fig:gradient}. Planets with low obliquities receive, on average, more stellar radiation at the equator than at the poles and usually feature a marked meridional temperature gradient. This situation is reversed at high obliquities (e.g. $\theta=90^\circ$) with polar regions receiving on average more radiation than equatorial regions.

\begin{figure*}[htb!]
\centering \includegraphics[trim = 5mm  15mm 10mm  130mm,angle=0,width=0.99\textwidth]{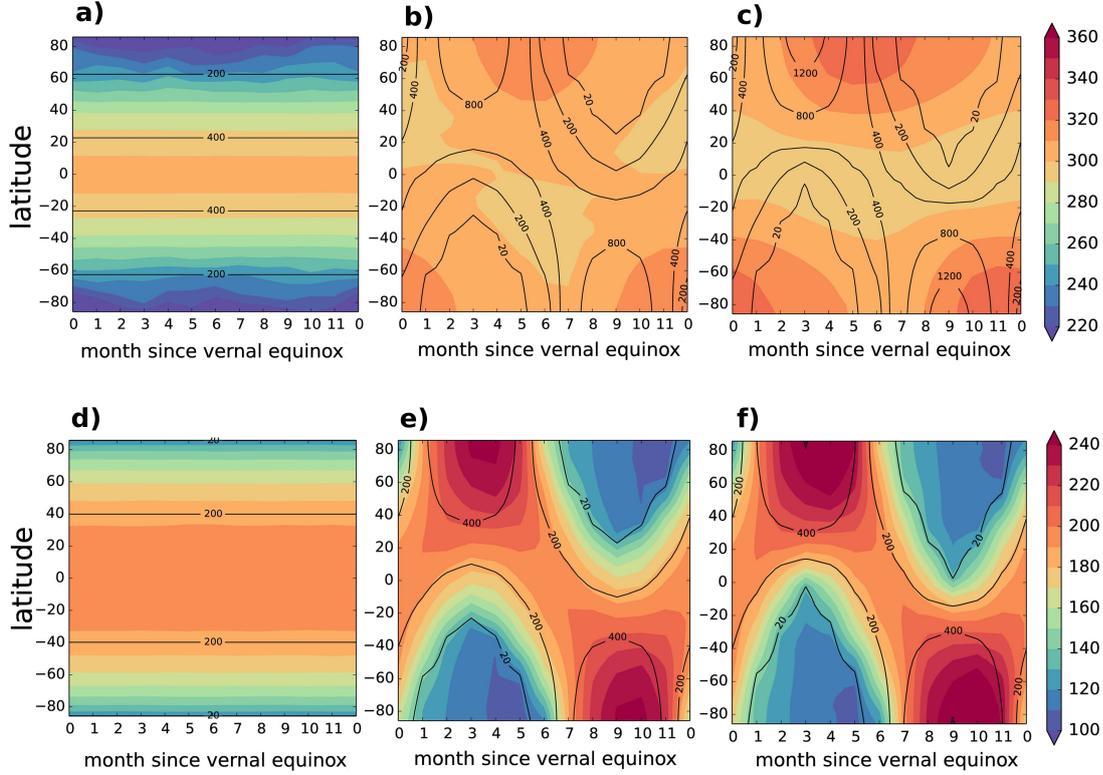}
\caption{Annual cycle of zonal mean incoming stellar radiation (contours, W\,m$^{-2}$) and surface temperature (shading, K)  of planets on a circular orbit ($e=0$); warm climate states (a) $\theta=0^\circ$, $\tilde{S}=1$, (b) $\theta=60^\circ$, $\tilde{S}=1$, (c) $\theta=90^\circ $, $\tilde{S}=1$; and snowball states ((d) $\theta=0^\circ$, $\tilde{S}=0.6$, (e) $\theta=60^\circ$, $\tilde{S}=0.6$, (f) $\theta=90^\circ$ , $\tilde{S}=0.6 $ on a circular orbit ($e=0$)} 
\label{fig:sol_for}
\end{figure*}

\begin{figure*}[htb!]
\centering \includegraphics[trim = 5mm  15mm 10mm  130mm,angle=0,width=0.99\textwidth]{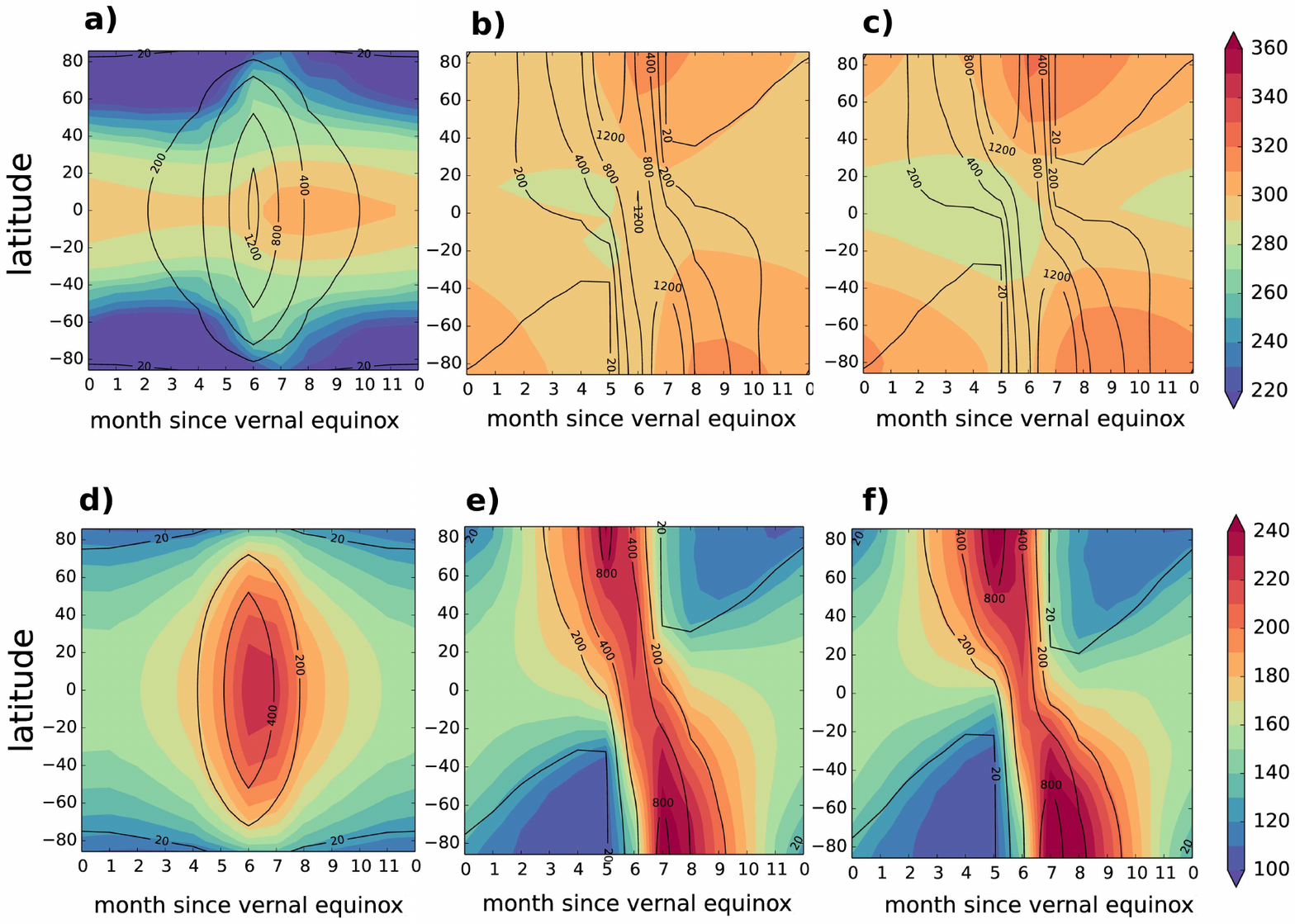} 
\caption{Annual cycle of zonal mean incoming stellar radiation (contours, W\,m$^{-2}$) and surface temperature (shading, K) as in Fig.~\ref{fig:sol_for} but of planets on an eccentric orbit with $e=0.5$; warm climate states (a) $\theta=0^\circ$, $\tilde{S}=0.8$, (b) $\theta=60^\circ$, $\tilde{S}=0.8$, (c) $\theta=90^\circ $, $\tilde{S}=0.8$; and snowball states ((d) $\theta=0^\circ$, $\tilde{S}=0.4$, (e) $\theta=60^\circ$, $\tilde{S}=0.4$, (f) $\theta=90^\circ$, $\tilde{S}=0.4 $}
\label{fig:sol_for2}
\end{figure*}

Large seasonal deviations from the annual mean temperature are typical of climates at high obliquity. Patterns of zonal-mean surface temperature and zonal-mean net incoming stellar radiation are shown for each month in Fig.~\ref{fig:sol_for} and Fig.~\ref{fig:sol_for2} for ice-free and snowball states for different values of obliquity and eccentricity. Because snowball states have low thermal inertia, the patterns of surface temperature follow closely the patterns of the net incoming stellar radiation. For ice-free climate states, the heat capacity of the slab ocean causes a delay of the temperature response to solar forcing of about two to three months.

The seasonal cycle of insolation at both high obliquity and eccentricity has a complex pattern, shifting the times of maximum insolation in the northern and southern hemisphere closer to each other (from month 3  to month 5 and from month 9 to month 7 respectively). Although solstice occurs on both poles at the same distance from the star (azimuthal obliquity $\omega=0^\circ$), there is a slight asymmetry in the hemispheric temperature patterns  at $\theta=60^\circ$ and $\theta=90^\circ$. This is due to the fact that in the northern hemisphere summer is followed by the planet getting closer to the star, while in the southern hemisphere summer is followed by the planet moving away from the star. That is, both poles receive the same overall amount of energy, but distributed differently over time, which due to the thermal inertia gives rise to  the asymmetric shape of surface temperature observed in Fig.~\ref{fig:sol_for} and ~\ref{fig:sol_for2}.

\begin{figure*}[htb]
\centering \includegraphics[trim = 10mm  5mm 20mm  130mm,angle=0,width=0.89\textwidth]{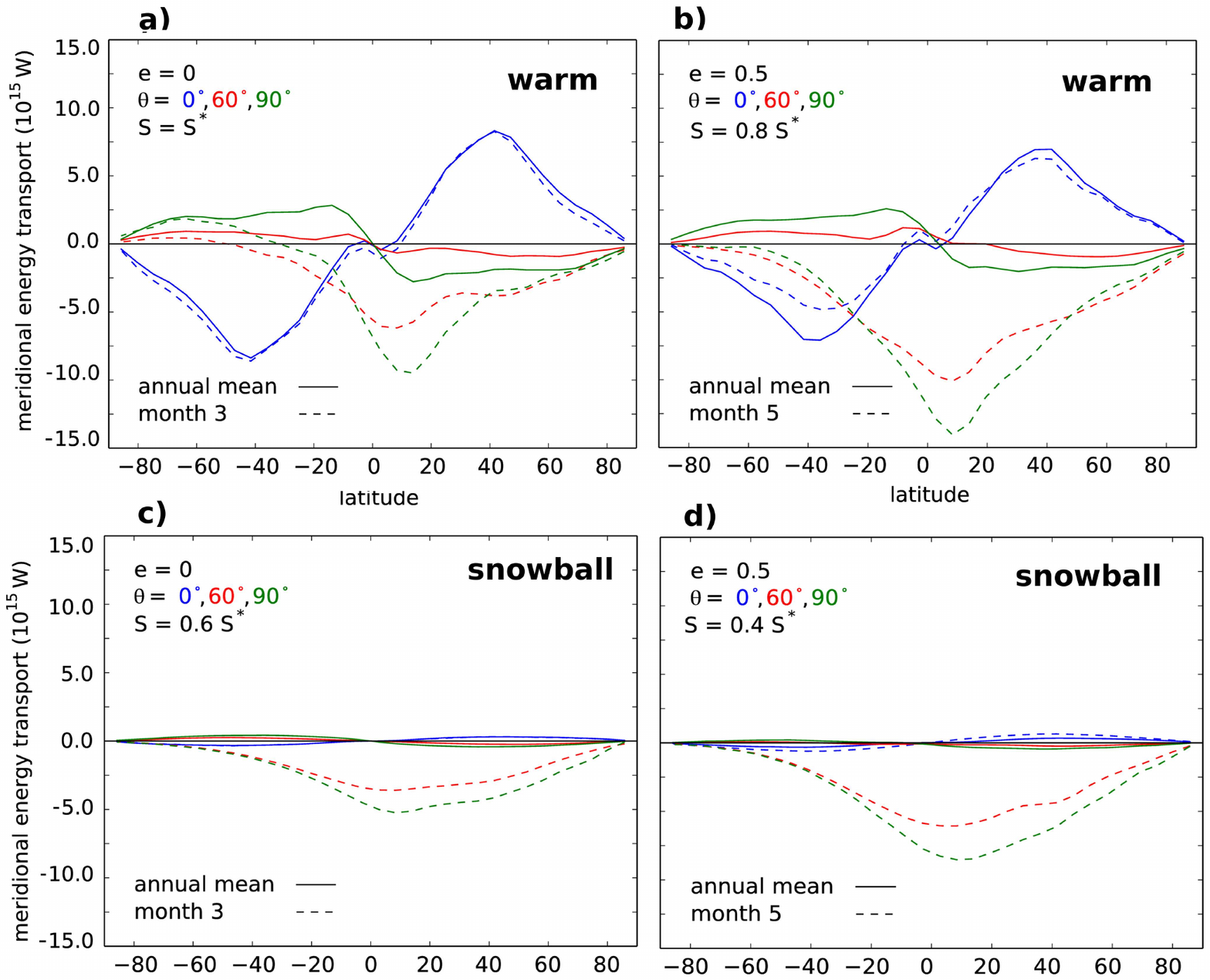}
\caption{Total meridional transport of moist static energy for (a) a warm climate state with a circular orbit, (b) a warm climate state with an eccentric orbit, (c) a snowball state with a circular orbit, and (d) a snowball state with an eccentric orbit; each for $\theta = 0^{\circ}$, $\theta = 60^{\circ}$ and  $\theta = 90^{\circ}$; months 3 and 7 correspond to the maximal inter-hemispheric difference in incoming solar radiation}
\label{fig:energy}
\end{figure*}

Horizontal heat transport is essential for smoothing the effect of extreme seasonal insolation at high obliquity. Since the net incoming stellar radiation features extreme seasonal variations, also the meridional energy transport shows extreme changes over one year (Fig.~\ref{fig:energy}). For planets at high obliquity (Fig.~\ref{fig:energy}(a)), the mean annual transport is equatorwards at $\theta = 90^\circ$ and nearly zero at $\theta = 60^\circ$. During the month of maximum irradiation in the northern hemisphere (Fig.~\ref{fig:sol_for}), the meridional heat transport is southwards at almost all latitudes, substantially differing from the annual mean values.  Deviations from the annual means become extreme for frozen planets (Fig.~\ref{fig:energy}(c)). Eccentricity  introduces only small variations to the annual mean meridional heat transport  but changes the timing of its seasonal oscillations because of shifted and distorted insolation patterns ($e=0.5$, Fig.~\ref{fig:sol_for}(b, d)).

The atmospheric meridional energy transports follows directly from the atmospheric circulation. Circulations of earth-like planets with low obliquity ($\theta\rightarrow 0$)  are similar to those of the Earth's atmosphere, with equatorial regions warmer than polar regions, a direct Hadley cell confined in the Tropics and jet streams at mid-latitudes \citep[e.g.][]{James}.  Snowball Earth-like climates for planets at low obliquity are characterized by a very dry atmosphere (neither the condensation latent heat nor the water vapor greenhouse effect have any significant impact on climate) and by very weak vertical gradients. These climates have intensively been discussed in the literature \citep[e.g.~][]{Pierrehumbert_2011b,Abbot_2013, Boschi_2013}.

\begin{figure*}[htb!]
\centering \includegraphics[trim = 0mm  10mm 60mm  90mm,angle=0,width=0.8\textwidth]{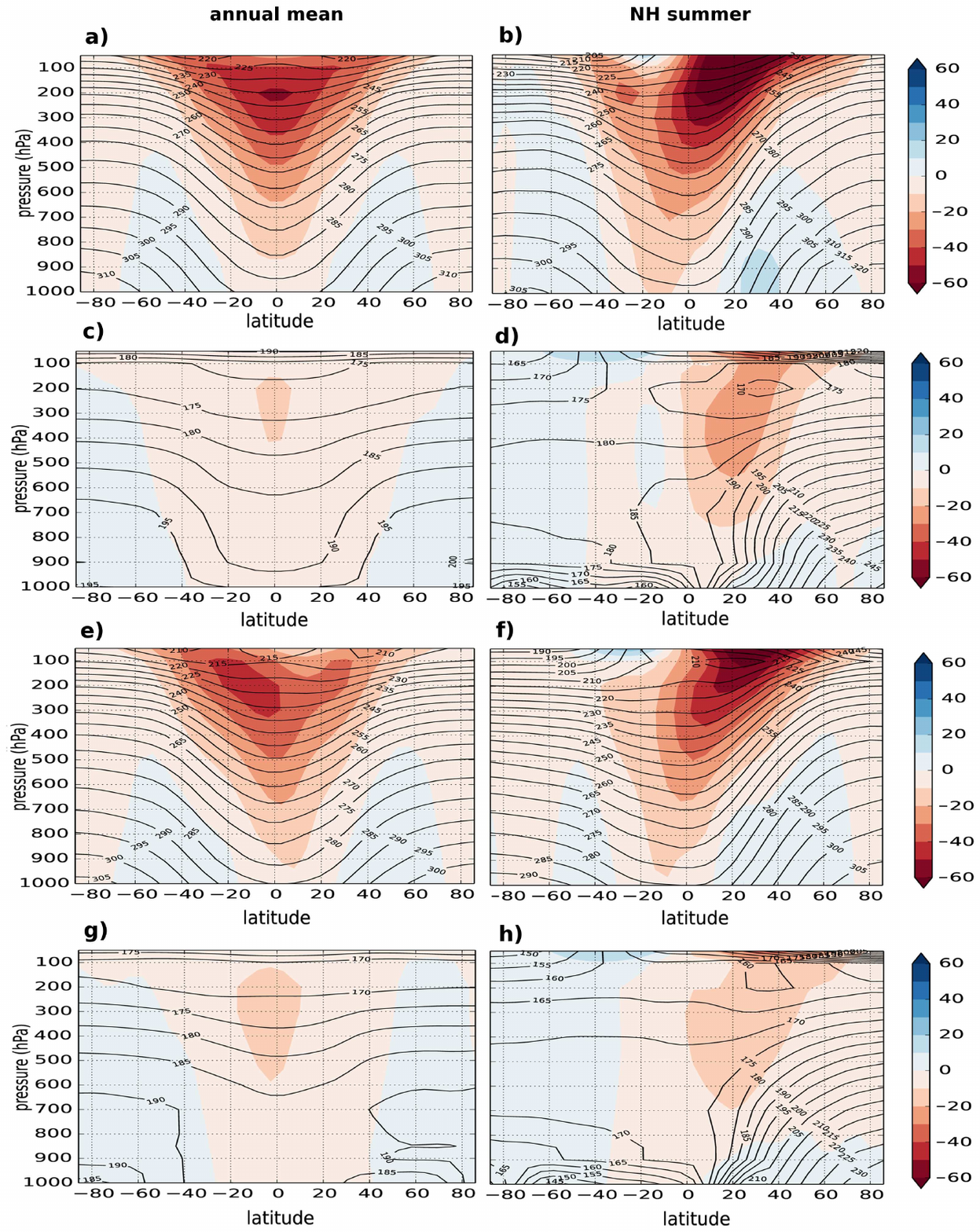}
\caption{Annual mean (left column) and monthly mean for summer in northern hemisphere (right column) of zonal winds (m\,s$^{-1}$) and atmospheric temperature (K); warm states (a, b) with $\theta=90^\circ$, $e=0$ ($\tilde{S}=1$); snowball states (c, d)  with $\theta=90^\circ$, $e=0$ ($\tilde{S}=0.6$); warm states (e, f) with  $\theta=90^\circ$, $e=0.5$ ($\tilde{S}=0.8$) and snowball states (g, h) with $\theta=90^\circ$, $e=0.5$ ($\tilde{S}=0.4$)}
\label{fig:u_T}
\end{figure*}

\begin{figure*}[htb!]
\centering \includegraphics[trim = 0mm  5mm 60mm  90mm,angle=0,width=0.8\textwidth]{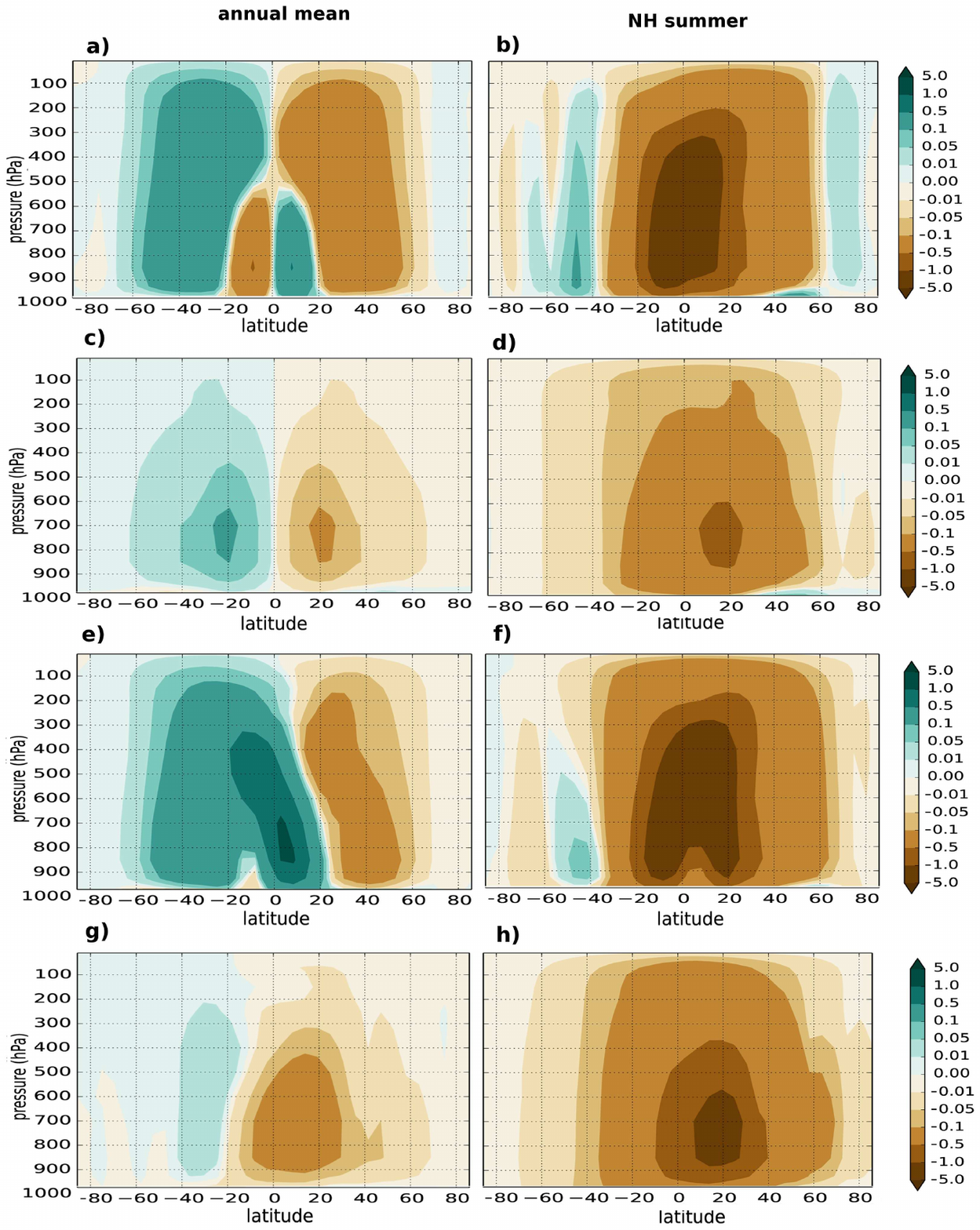}
\caption{Annual means (left column) and northern hemisphere (right column) of  the Eulerian mean stream function (in $10^{11}$\, kg\,s$^{-1}$); warm states (a, b) with $\theta=90^\circ$, $e=0$ ($\tilde{S}=1$); snowball states (c, d) with $\theta=90^\circ$, $e=0$ ($\tilde{S}=0.6$); warm states (e, f) with $\theta=90^\circ$, $e=0.5$ ($\tilde{S}=0.8$); snowball states (g, h) with $\theta=90^\circ$, $e=0.5$ ($\tilde{S}=0.4$); the circulation goes clockwise along positive streamlines}
\label{fig:stream}
\end{figure*}

High obliquity causes a reversal of the atmospheric meridional temperature profile with respect to climates at low obliquity, because on annual mean equatorial regions receive more solar radiation than polar regions (Fig.~\ref{fig:u_T}(a, c, e, f)). In Fig.~\ref{fig:u_T}(a) the annual means of atmospheric temperatures and zonal winds are shown. The meridional streamfunction can be see in  in Fig.~\ref{fig:stream}(a) for the same temporal means. Since the circulations show pronounced seasonal variations, also the warmest month in the northern hemisphere during which there is the strongest meridional temperature contrast is included in Fig.~\ref{fig:stream}.

Westerly storm-tracks develop in the summer hemisphere in the lower troposphere, with a strongly sheared easterly jet aloft for the warm climate at high obliquity (FIg.~\ref{fig:u_T}(b). The atmospheric meridional overturning circulation during the NH summer (Fig.~\ref{fig:stream}(b)) is thermally direct (upwelling in the northern hemisphere and downwelling in the southern  hemisphere) and extends over the whole hemisphere. The reversal of the direct cell and its shift in the southern hemisphere during the local summer results in the two small equatorial indirect cells seen in the annual mean (Fig.~\ref{fig:stream}(a)), which hence are just an artefact of the annual averaging. In the winter hemisphere no strong westerly jet develops close the ground.  \citet{Ferreira_2014} show that  the strongly sheared easterly winds in the mid-high troposphere are essential to determine baroclinic instability leading to the development of the westerly jet close to the ground, which are missing in the southern/winter hemisphere. Circulations of Earth-like planets at high obliquity and on eccentric orbits differ only slightly from the pattern discussed so far. The main difference is the asymmetric circulation between the two hemispheres (Fig.~\ref{fig:u_T}(e, f) and Fig.~\ref{fig:stream}(e, f)) reflecting the asymmetric temperature patterns already discussed and shown in Fig.~\ref{fig:sol_for}(c) and Fig. ~\ref{fig:sol_for2}(c).

One of  main features of snowball climates is the replacement of the ocean (in our simulations represented by a slab-ocean model) with a thick ice cover, which drastically reduces the thermal inertia of the ocean. As a consequence, surface temperatures respond almost instantaneously to the distribution of absorbed solar radiation (Fig.~\ref{fig:sol_for}, ~\ref{fig:sol_for2}).  As for snowball climates at low obliquity (for which a rich literature exists, \citep[e.g.~]{Pierrehumbert_2011b, Abbot_2013, Boschi_2013}) snowball climates at high obliquity feature very weak vertical thermal gradients (Fig.~\ref{fig:u_T}(c, g)). This is a direct consequence of the dryness of the atmosphere and therefore  the lack of greenhouse effect. Important deviations from the annually-averaged temperature profile are however observed seasonally. Fig.~\ref{fig:u_T}(d, h) refers to  the northern hemisphere summer and shows a  very stable temperature profile in the southern hemisphere, with an atmosphere which is  almost isothermal,  except close to the surface where a strong thermal inversion is found. The vertical profile is instead destabilized in the summer hemisphere by the intense irradiation at the ground. Air rises in the summer hemisphere forming a thermally direct cell which extends globally up to the south pole (Fig.~\ref{fig:stream}(d, h)).  The solstitial circulation patterns dominate the annal mean,  resulting in a fictitious two-cell structure of the annually-averaged meridional streamfunction (Fig.~\ref{fig:stream}(c, g)). As already noted above, changes in insolation due to eccentricity do not substantially impact the main circulation features of snowball climates at high obliquity, but introduce an asymmetric behaviour between the northern and southern hemisphere (Fig.~\ref{fig:stream}(g)).

\section{Multistability and habitability at extreme seasonality}\label{sec:results}

Within a certain range of $\tilde{S}$, two distinct climate states coexist, one warm (mostly ice-free) and one cold (mostly ice-covered) climate state \citep{Budyko_1969,Sellers_1969,Ghil_1976,Bodai_2014}. Building on previous work on the bistability and habitability of Earth-like planets \citep{Boschi_2013,Lucarini_2013} the implications of seasonal variability on the stable intervals of both climate states are examined in the following. For each set of parameters, a hysteretic cycle is performed. We start the model at an irradiation around $\tilde{S}=1$ and then gradually decrease it by steps of $0.05$ until the planet is completely covered with ice. For each value of $\tilde{S}$, the model is run for fifty years, of which the last thirty years are used for the computation of climatological quantities. Once the planet is completely ice-covered, $\tilde{S}$ is step-wise increased again until the planet is completely ice-free. 

\subsection{Effect of obliquity on ice formation and melting} \label{sec:results:freeze} 

We begin our analysis by investigating the effect of obliquity on ice-formation and ice-melt on circular orbits, where the obliquity determines both the spatial distribution of insolation and its temporal variability. Our results show that both the mean distribution and its variability have substantial consequences for the transitions between the completely ice-free and the completely ice-covered climate state.

\begin{figure*}[htb!]
\centering
\includegraphics[width=0.49\textwidth]{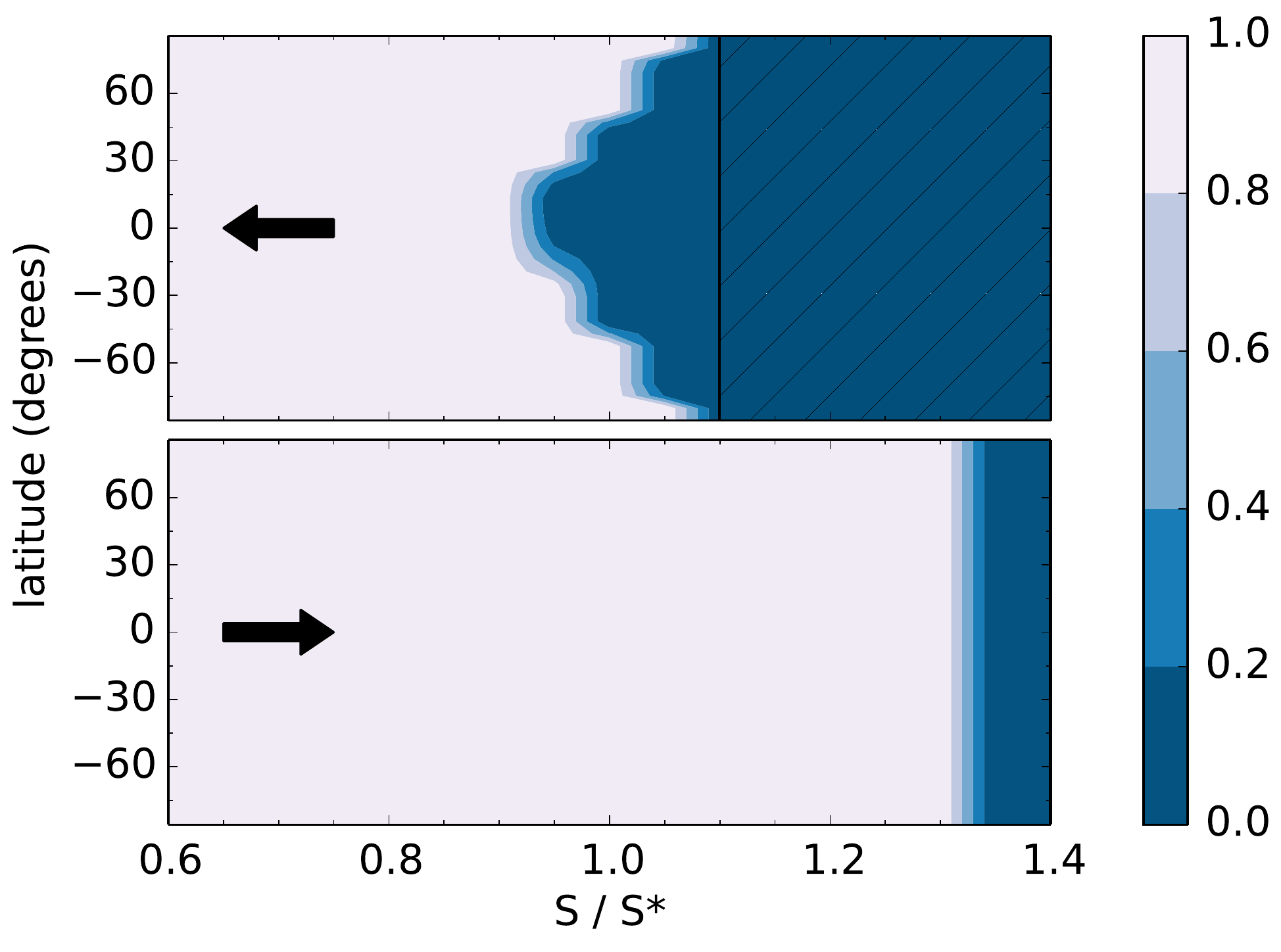}
\includegraphics[width=0.49\textwidth]{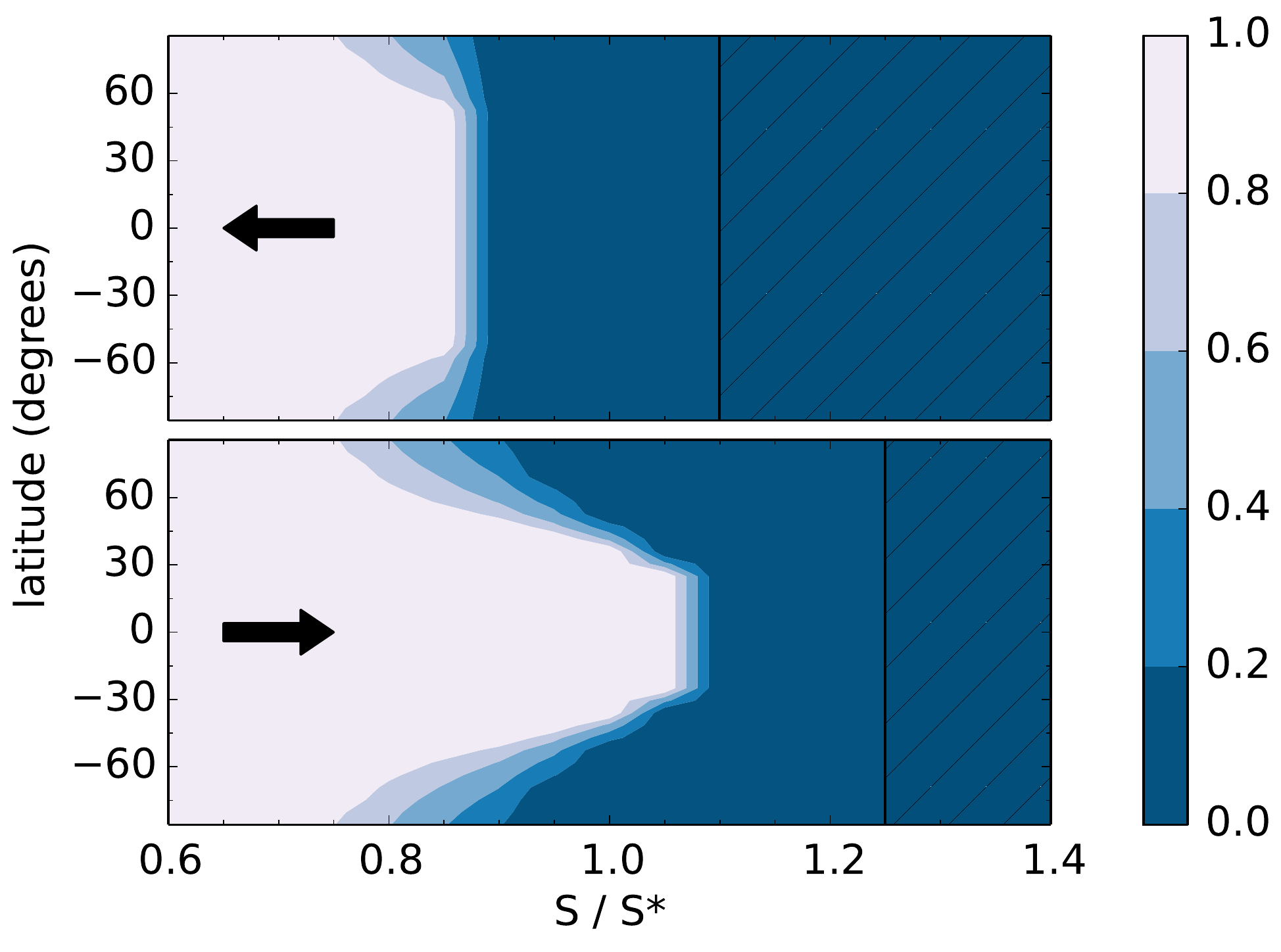}
\caption{Ice coverage (as fraction of the year) for a planet on a circular orbit with an obliquity of $\theta = 0^{\circ}$ (left) and $\theta = 90^{\circ}$ (right); the zonal means of the last 10 years of each steady state are shown; the arrows indicate the direction of change of $\tilde{S}$, i.e. whether an ice-free or ice-covered initial state is assumed; hashes indicate areas outside the simulated range of $\tilde{S}$}
\label{fig:trans}
\end{figure*}

Planets with low obliquity receive most radiation at the equator, whereas planets with high obliquity receive most of their energy at the poles (Figure \ref{fig:ins}). Consequently, when decreasing $\tilde{S}$, ice formation starts at the poles and the equator for $\theta = \SI{0}{\degree}$ and $\theta = \SI{90}{\degree}$, respectively (Figure \ref{fig:trans}). This can be correlated with the spatial pattern of surface temperature (not shown) that roughly follows the distribution of annual mean insolation (Figure \ref{fig:ins}).

Planets with low obliquity are more susceptible to global freezing than planets with high obliquity (Figure \ref{fig:trans}). On planets with low values of $\theta$, strong gradients of meridional temperature exist that can allow for the existence of ice at the poles despite relatively high temperatures at the equator (Figure \ref{fig:gradient}). Once ice forms at the poles at a certain value of $\tilde{S}$, a further decrease of $\tilde{S}$ leads to an advancement of the ice edge towards the equator. This leads to a higher planetary albedo and thus a further cooling of the planet, which finally results in a global freezing at relatively high values of $\tilde{S}$. In contrast, temperature distributions on planets with high values of $\theta$ are more homogeneous (Figure \ref{fig:gradient}). This implies that ice does not form for relatively low values of $\tilde{S}$ and hence the ice-albedo feedback does not amplify the cooling due to less incoming radiation. Yet once a critical threshold of $\tilde{S}$ is crossed and the temperature falls below the freezing point at the equator, a rather sharp transition to a completely frozen state takes place. Consequently, the transition to the completely frozen state occurs in a more gradual manner and for a higher value of $\tilde{S}$ on the planet with $\theta = \SI{0}{\degree}$ than for $\theta = \SI{90}{\degree}$ (Figure \ref{fig:trans}).

The dynamics of ice-formation and ice-melt is reversed for planets with high and low obliquity, because the geography of ice is reversed in the two cases. Melting the ice sheet of planets with $\theta = \SI{0}{\degree}$ requires relatively high values of $\tilde{S}$, as seasonal peak irradiation is not strong enough to trigger an initial appearance of ice-free areas at the equator. Once the critical threshold is crossed and initial ice-free areas appear, the ice-albedo feedback leads to a sharp transition. As opposed to this, planets with high values of $\theta$ feature ice-free areas at the poles already at lower values of $\tilde{S}$. Hence also the transition to a completely ice-free state occurs in a more gradual manner (Figure \ref{fig:trans}).

\subsection{Effect of obliquity and eccentricity on multistability}

We compute the global mean surface temperature for different values of irradiation $\tilde{S}$ in order to summarize information about the climate state into one observable. The choice of surface temperature as observable is motivated by two aspects. First, the observable can directly be linked to usual habitability conditions requiring the surface temperature to remain between the freezing and boiling point of water. Second, the observable depends strongly enough on ice-coverage to show a clear shift at the point of transition between the two distinct climate states. Also alternative ways to compress information about the climate state in one quantity based on thermodynamics have been proposed and applied in similar studies, revealing a good correlation between global mean surface temperature, material entropy production and quantities indicative of the efficiency of the climate machine \citep{Lucarini_2010,Boschi_2013}. However, because our focus is on average climate conditions, we choose surface temperature as observable in our analysis of climate stability. In Section \ref{sec:results:hab} we develop a further index that is more relevant in the discussion about habitability.

\begin{figure*}[htb!]\centering 
\includegraphics[width=0.8\textwidth]{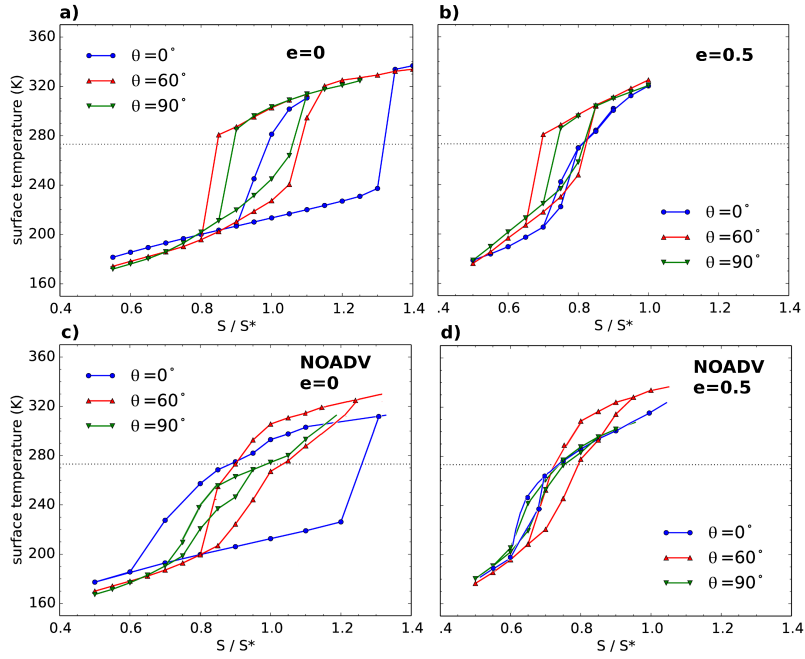}
\caption{Annual and global mean of surface temperature versus normalized stellar irradiation for planets with obliquity $\theta = 0^{\circ}, 60^{\circ}$ and $90^{\circ}$ and eccentricity (a) $e = 0$ and (b) $ e = 0.5$; (c, d) hysteresis curves obtained by running the same simulations but without any horizontal transport processes (NOADV)}
\label{fig:bifurc}
\end{figure*}

In all experiments, the relationship between global mean surface temperature and solar irradiation exhibits two branches over a certain range of $\tilde{S}$. Over this range of $\tilde{S}$, the actual climate state depends on the history of the planet (Figure \ref{fig:bifurc}(a, b)). This phenomenon is known as climate bistability, or multistability. Where both climate states exist we distinguish them as warm and cold (snowball) climate state.

As explained above, the more homogeneous the distribution of surface temperature (Figure \ref{fig:gradient}), the lower the critical value of $\tilde{S}$ at which the planet enters the cold state. Here, an obliquity $\theta = \SI{60}{\degree}$ shows the largest extent of the warm state, followed by $\theta = \SI{90}{\degree}$ (Figure \ref{fig:bifurc}(a, b)). In the cold state, ice-free areas appear once seasonal insolation at any latitude is intense enough to melt the ice cap. Therefore, the extent of the cold state is minimal for high values of obliquity (Figure \ref{fig:bifurc}), which are associated with very high seasonal peak irradiation.

The effect of obliquity on the extent of the warm state is qualitatively the same for a circular orbit as for an orbit with $e=0.5$. Also on such orbit, the largest extent of the warm state is shown by the planet with $\theta = 60^{\circ}$, followed by $\theta = 90^{\circ}$ (Figure \ref{fig:bifurc}(a, b)). However, the range of bistability shrinks for all experiments with $e=0.5$ compared to the experiments with $e=0$. Moreover, all experiments with $e=0.5$ show a larger extent of the warm state than the corresponding experiments with $e=0$ (Figure \ref{fig:bifurc}(a, b)). This can in part be explained with the increase of the global and annual mean insolation $\langle I \rangle_{e}$ on an orbit with eccentricity $e$ \citep{Dressing_2010}, whose effect is investigated separately below:

\begin{equation} \langle I \rangle _{e} = \frac{\langle I \rangle _{e=0}}{\sqrt{1 - e^2}}. \label{eq:ins_ecc} \end{equation}

Also the extent of the cold state is strongly affected by the eccentricity of the orbit. In particular for $\theta= 0^{\circ}$, the width of the cold state is much smaller for $e=0.5$ than for $e=0$ (Figure \ref{fig:bifurc}(a, b)). This is due to the intense irradiation around periastron. While the irradiation on a planet with $\theta= 0^{\circ}$ and $e=0$ is not strong enough to lead to ice-free regions, the generally higher annual mean irradiation and the high peak irradiation around periastron on an orbit with $e=0.5$ suffice to melt the ice layer. This generally leads to a sharper transition between the two states and possibly only one stable state for higher eccentricities (Figure \ref{fig:bifurc}(a, b)). This bifurcation from two stable states to only one stable state was also observed in previous studies where the length of one year was varied \citep{Lucarini_2013}.

Meridional transport processes due to atmospheric dynamics play a crucial role in determining the behaviour of the ice-albedo feedback and therefore also affect the transitions between ice-free and ice-covered climate states. Hysteresis curves shown in Fig.~\ref{fig:bifurc}(c, d) are obtained after ``switching off'' the atmospheric dynamic core of PlaSim, leaving each atmospheric column isolated from its neighbouring columns and in radiative-convective equilibrium. Because temperature differences are not smoothed by transport processes, large gradients build up between regions with more and less intense insolation. These simulations show that the absence of horizontal transport processes has as a consequence for the disappearance of the catastrophic events in the hysteresis curves. For different obliquities and different eccentricities the planets get gradually frozen without the sudden glaciation observed in full-dynamics simulations (Fig.~\ref{fig:bifurc}(c, d)). This is in line with the analysis of results obtained from one-dimensional EBM of Earth, showing that the sensitivity of the ice-line to changes in solar radiation, which goes to infinity during a snowball catastrophe, is identically zero if there is no meridional heat transport \citep{Roe_2010}. As concluding remark we note that, although catastrophic events represented by sudden jumps of the mean surface temperature disappear when transport processes are excluded, hysteresis still remains with curves showing characteristics different between the two directions of the transition.

\subsection{Implications for habitability}\label{sec:results:hab}

The existence of liquid water is the most common criterion for habitability. In estimates of the outer boundary of the habitable zone with zero dimensional EBMs or RCMs, this has been translated to the requirement of a global average surface temperature above the freezing point of water. Another sufficient condition for habitability is the existence of a warm stable climate state, because in the case of bistability the warm state always features some ice-free areas. These two conditions are however conservative and neglect that in case of spatial and temporal heterogeneity ice-free areas can also exist under both conditions, at a global mean surface temperature below the freezing point and in a cold climate state.

We therefore use a slightly modified version of the definition of temporal habitability proposed by \citet{Spiegel_2008}, based on sea ice cover instead of surface temperature as measure for habitability. This reflects our use of an explicit sea ice model and the fact that stability issues of PlaSim do not allow to simulate climate close to the boiling point of water, which restricts our analysis to the outer boundary of the HZ. We then compute the zonal habitability $H(\varphi, t)$ from sea ice cover $\sigma(\varphi, \lambda, t)$ according to

\begin{equation} H(\varphi, t) = 1 - \frac{1}{2\pi}\int_{0}^{2\pi} \sigma(\varphi, \lambda, t) \rm{d}\lambda, \end{equation}

where $\varphi$ and $\lambda$ are latitude and longitude respectively. The temporal habitability $f(\varphi)$ is then defined as

\begin{equation} f(\varphi) \; = \; \frac{\int_{t = 0}^{T}\;H(\varphi, t)\; dt}{T} \end{equation}

where $T$ denotes the length of one year. We then classify each climate state as one of the following: \begin{itemize}
\item \textit{habitable} if $f(\varphi) = 1$ at all latitudes $\varphi$
\item \textit{partially habitable} if $f(\varphi) = 1$ at some latitude
\item \textit{temporally habitable} if $f(\varphi) > 0$ at some latitude but
$f(\varphi) < 1$ at all latitudes
\item \textit{unhabitable} if $f(\varphi) = 0$ at all latitudes \end{itemize}

\begin{sidewaysfigure*}[htb!]
\centering \includegraphics[width=0.3\textwidth]{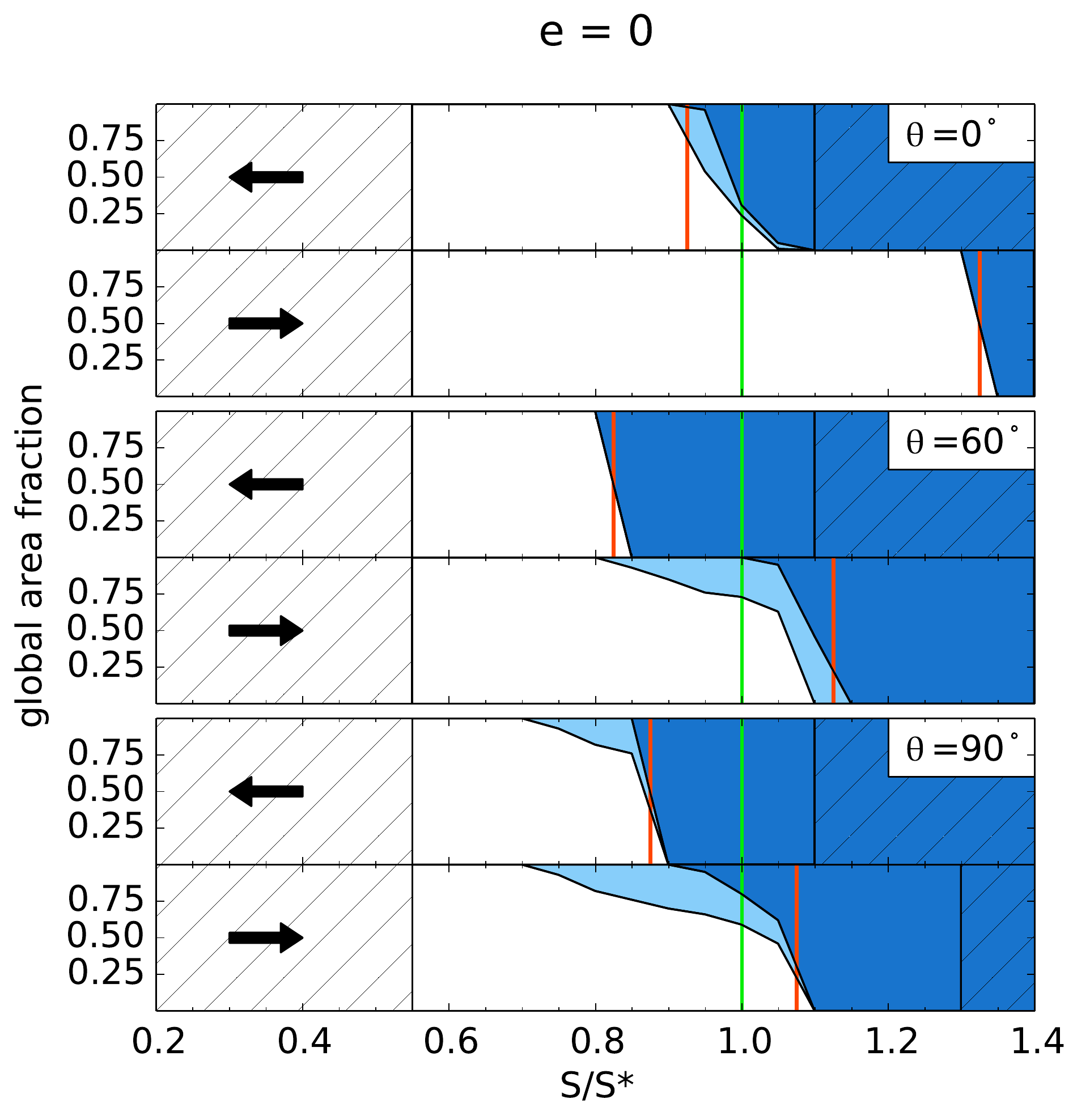}
\includegraphics[width=0.3\textwidth]{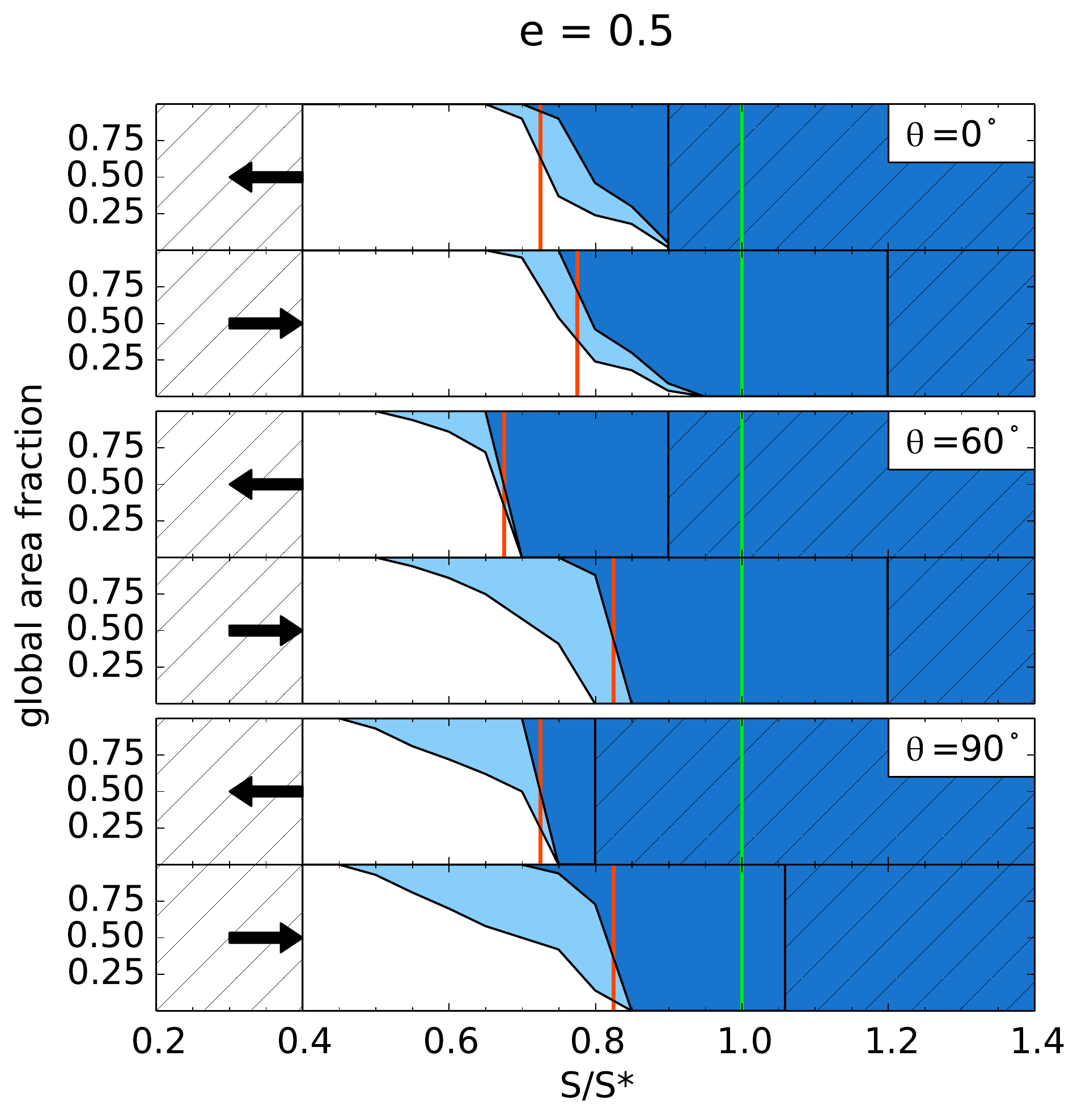}
\includegraphics[width=0.3\textwidth]{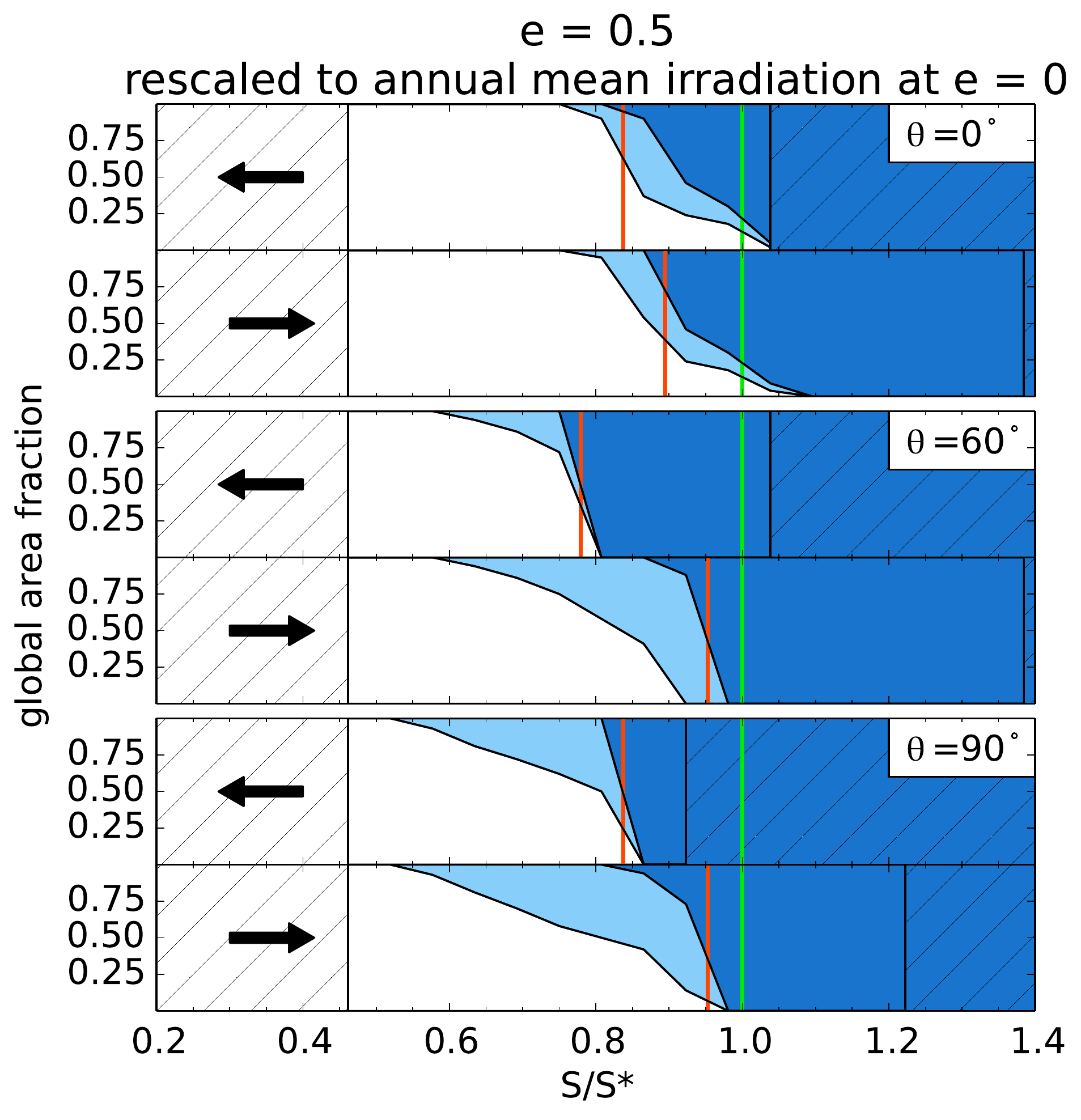}
\caption{Degree of habitability for simulations with $e = 0$ (left) and $e = 0.5$ (center), and $\theta = 0, 60, 90^{\circ}$ (from top to bottom); the color denotes the fraction of the planet's surface that is (partially) habitable (dark blue), temporally habitable (light blue) and unhabitable (white); the red vertical line marks the point of transition between the warm and cold state; the solid green vertical line shows the solar irradiation at $\SI{1}{AU}$ as reference; (right) results for $e=0.5$ rescaled according to the annual mean irradiation $\langle I\rangle _{e=0.5}$ (see Equation \ref{eq:ins_ecc}), such that the total annual amount of energy received from the star is the same as for $e= 0$ (left); the arrows indicate the direction of change of $\tilde{S}$, i.e. whether an ice-free or ice-covered initial state is assumed; hashes indicate areas outside the simulated range of $\tilde{S}$}
\label{fig:hab_degree}
\end{sidewaysfigure*}

Our experiments reveal several aspects of the effect of obliquity and eccentricity on a planet's habitability. First of all, seasonal variability can lead to a substantial extension of the habitable zone. While a planet with $e=0$ and $\theta = 0^{\circ}$ is habitable for $\tilde{S} > 0.95$ (warm initial state) or $\tilde{S} > 1.3$ (cold initial state), a planet with $e=0.5$ and $\theta = 90^{\circ}$ is habitable for $\tilde{S} > 0.55$ (E05o90, Figure \ref{fig:hab_degree}). Moreover, also cold climate states can exhibit regions that are temporally and even continously ice free. While the extent of the warm state is an appropriate measure of the HZ on planets without seasonal variations (E00o00, Figure \ref{fig:hab_degree}), it leads to a substantial underestimation of the HZ on planets with large seasonal variations (E05o90, Figure \ref{fig:hab_degree}).

On such planets the extension of the HZ is mainly due to regions that are temporally habitable, i.e ice free only at some time of the year. A clear distinction between partial and temporal habitability is therefore not only motivated from a biological perspective but also strongly suggested by the results of our climate simulations. Because eccentricity affects both the annual mean irradiation and its variability, these two effects have to be analysed separately. This is illustrated in the right column of Figure \ref{fig:hab_degree}. Here the results of the simulations at $e=0.5$ are rescaled to the mean irradiation at $e=0$ by dividing them through the factor at which the annual mean irradiation increases from $e=0$ to $e=0.5$ (see Equation 2). The results show that while the increase of annual mean irradiation with eccentricity is a good proxy for the extent of the warm state on an eccentric orbit, the simple scaling does neither capture the effects that lead to the extension of the HZ nor the effects that limit the extent of the cold state on an eccentric orbit. The remaining effect of eccentricity has to be attributed to the higher amplitude of the seasonal cycle and the different spatial and temporal distribution of irradiation (Figure \ref{fig:hab_degree}).

\begin{figure*}[htb!]
\centering \includegraphics[width=0.49\textwidth]{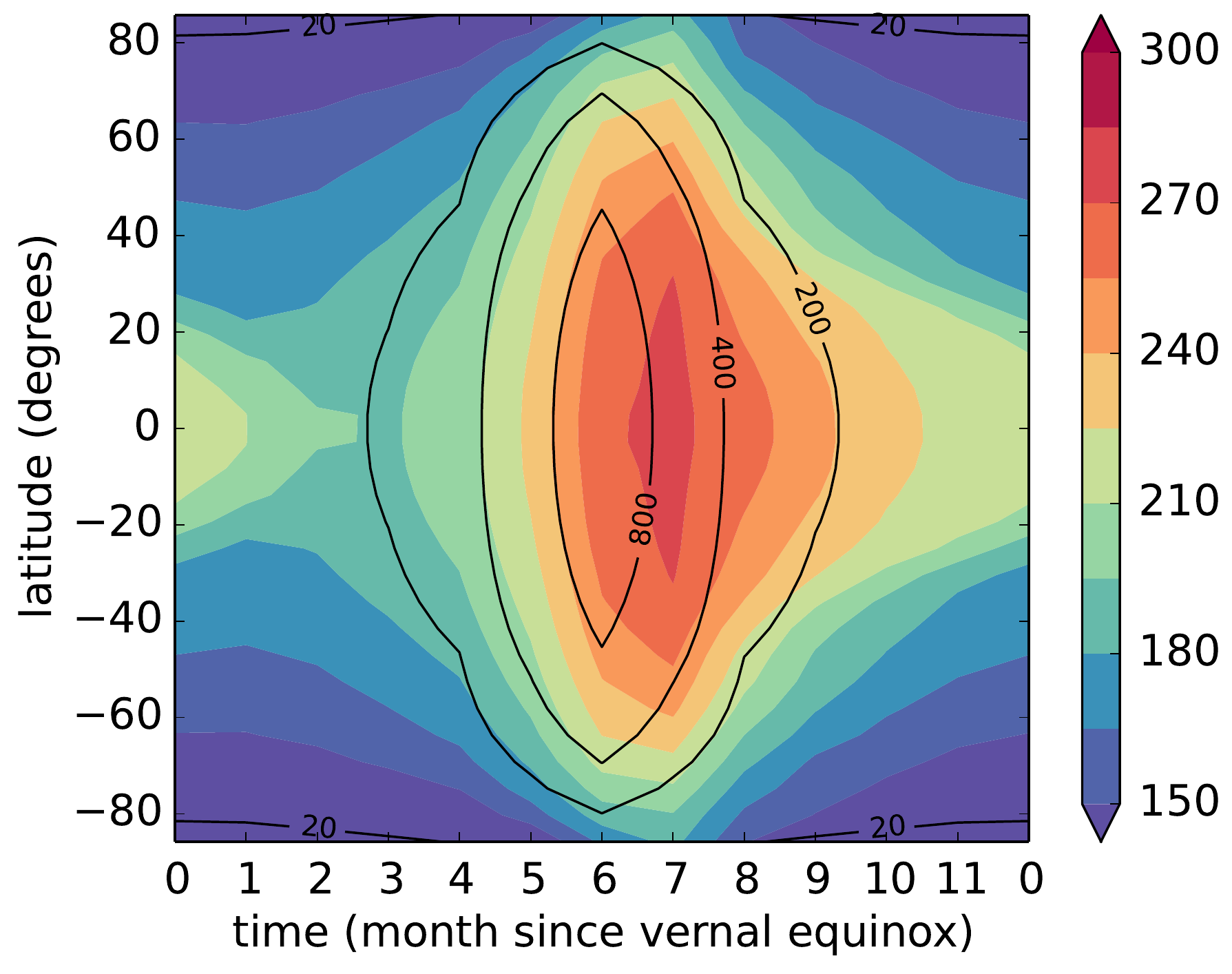}
\includegraphics[width=0.49\textwidth]{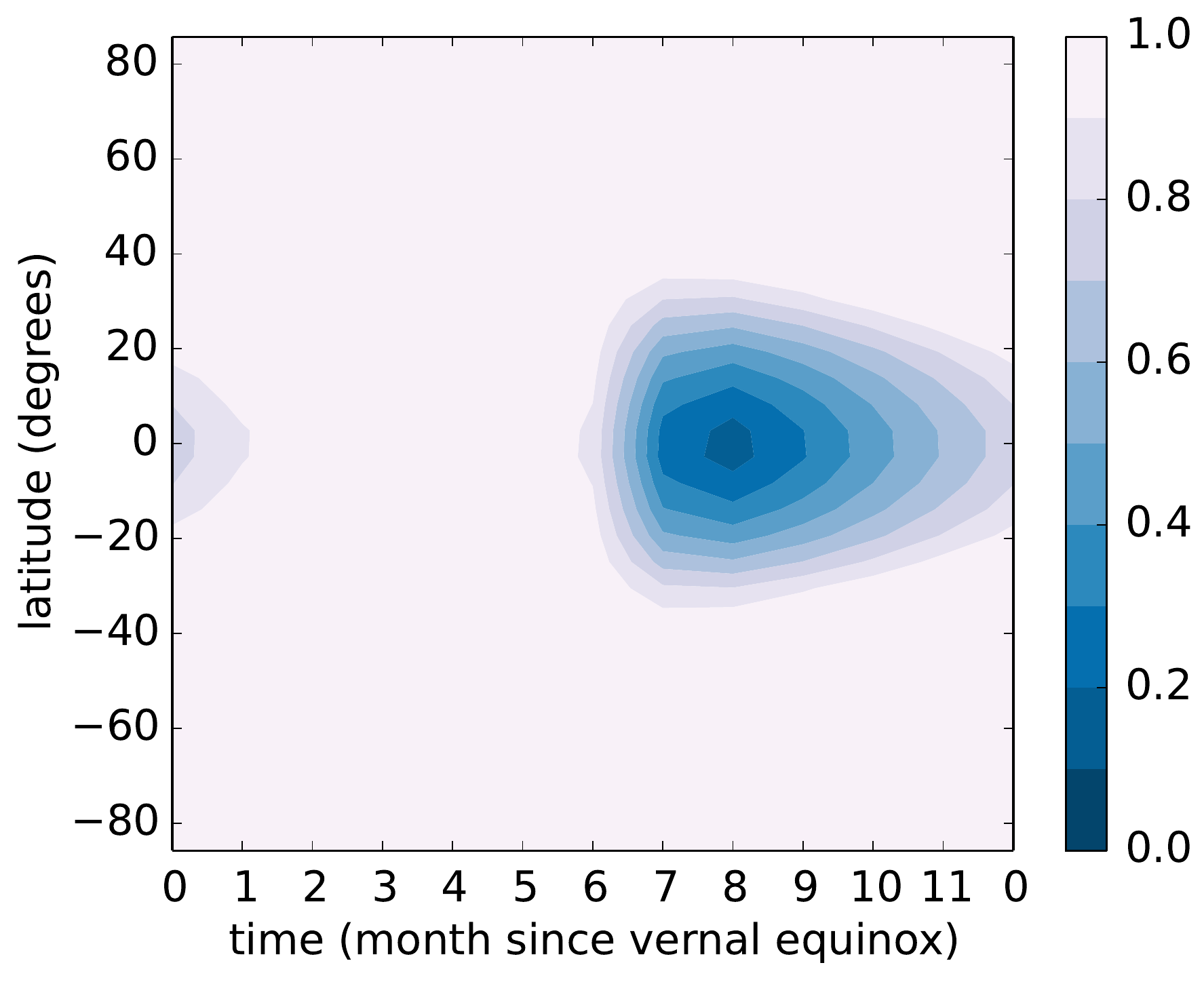}
\caption{Seasonal cycle of surface temperature (K; colors, left), irradiation $S$ (W/m$^{2}$; lines, left), and sea ice thickness ($h_{i}~/h_{i,\rm{max}}$; right) of the experiment E05o00 at $\tilde{S} = 0.7$ ($S = \SI{956}{Wm^{-2}}$)}
\label{fig:contour:00} 
\end{figure*}

\begin{figure*}[htb!]
\centering \includegraphics[width=0.49\textwidth]{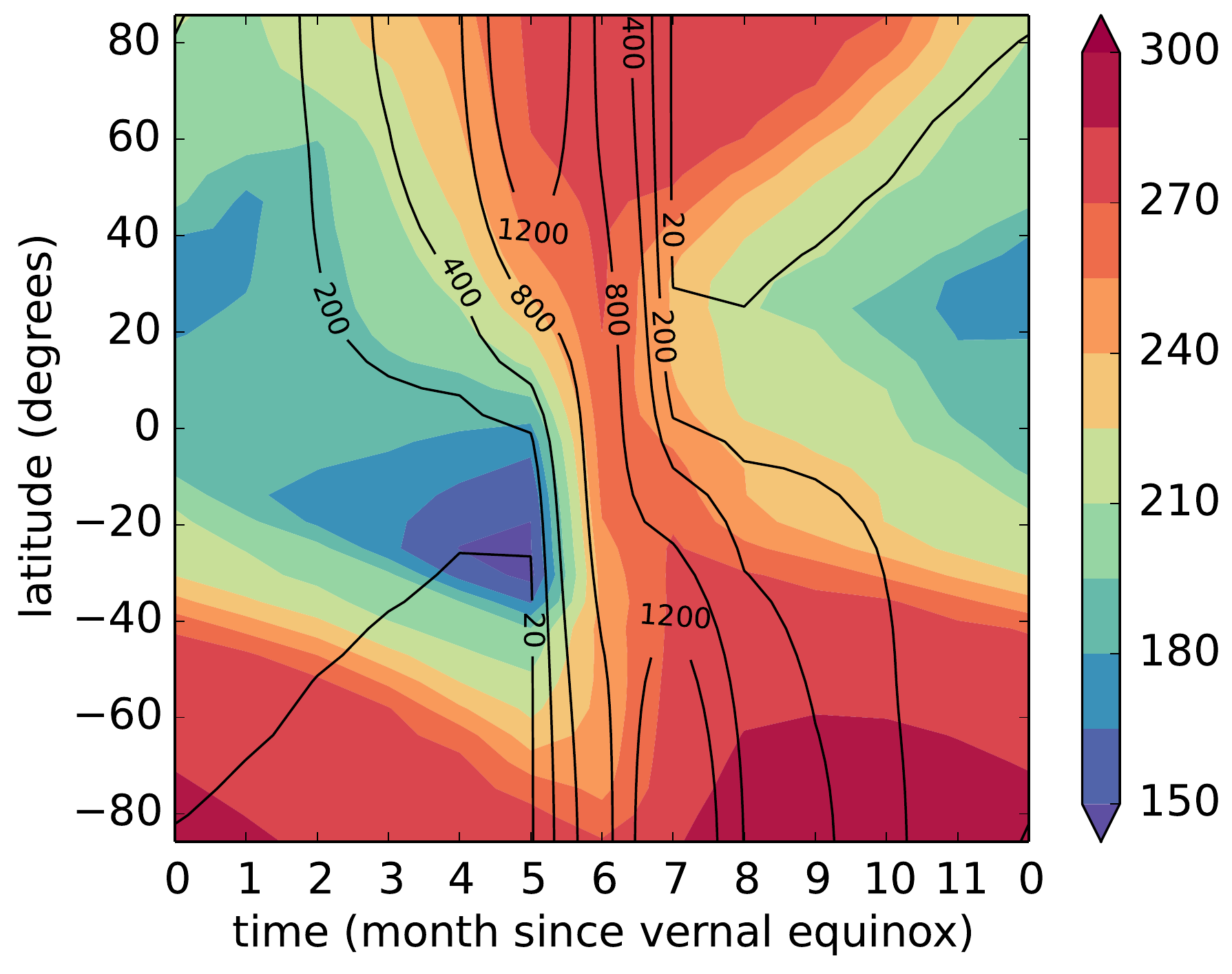}
\includegraphics[width=0.49\textwidth]{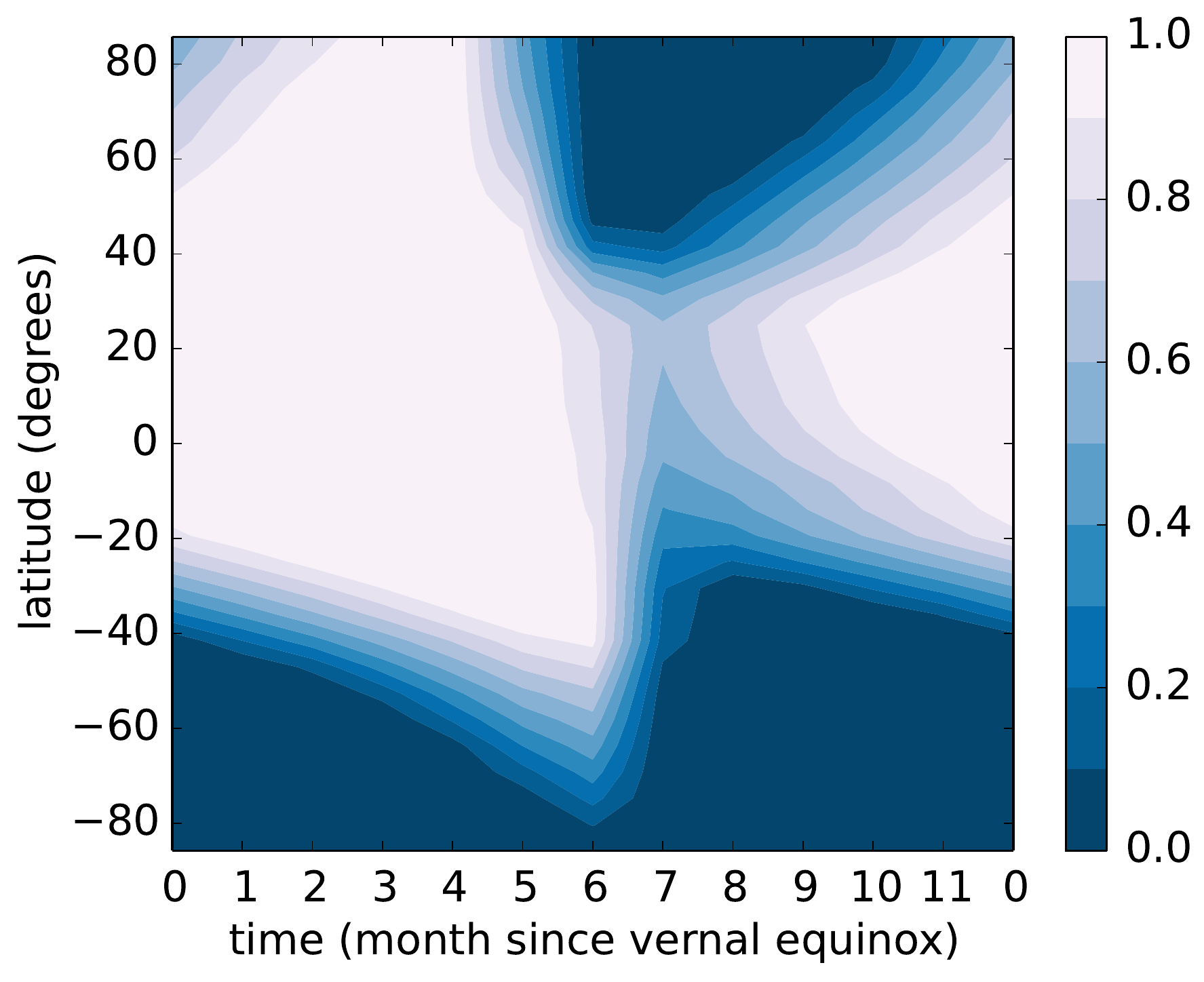} 
\caption{Seasonal cycle of surface temperature (K; colors, left), irradiation $S$ (W/m$^{2}$; lines, left), and sea ice thickness ($h_{i}~/h_{i,\rm{max}}$; right) of the experiment E05o90 at $\tilde{S} = 0.7$ ($S = \SI{956}{Wm^{-2}}$); note that the temporal evolution of $S$ differs among the two hemispheres due to the eccentric orbit, leading to an asymmetric distribution of surface temperature and ice-thickness}
\label{fig:contour:90}
\end{figure*}

Whether temporally habitable regions might indeed be suitable for life might depend on the amplitude of these seasonal variations and how long such regions might be ice-covered and ice-free. The presence of land masses (not accounted for in this study) may in turn facilitate or go against this partially habitable, ice-free regions. As illustrative examples we show the seasonal cycle of surface temperature and ice thickness of two simulations on an orbit with $e=0.5$ and $\theta = 0^{\circ}$ and $\theta = 90^{\circ}$. Both are shown for $\tilde{S} = 0.7$ (Figure \ref{fig:contour:00} and \ref{fig:contour:90}). The planet with $\theta = 0^{\circ}$ shows largest seasonal variations at low latitudes. Here surface temperatures exceed the freezing point for a short period around periastron, but long enough for a tiny area to become temporally ice free (Figure \ref{fig:contour:00}).

As compared to this, the climate of the planet with $\theta = 90^{\circ}$ is extremely variable, with almost half of the planet's surface experiencing freezing and complete melting in one year (Figure \ref{fig:contour:90}). Also surface temperatures show considerable oscillations with a maximal amplitude of the annual cylce of about 100 K at the equator. Apart from these unhabitable regions, however, high latitudes around the South Pole show rather moderate climatic conditions with a temperature oscillation of only about 20 K (Figure \ref{fig:contour:90}). 

The experiments shown in Figure \ref{fig:contour:00} and Figure \ref{fig:contour:90} differ only in the polar obliquity $\theta$. As both planets exhibit ice-free areas at some point of the year, they are equally classified as temporally habitable for $\tilde{S} = 0.7$. Yet a closer look reveals substantial differences with respect to the size of the temporally habitable area, the time this area is ice-free, and the amplitude of surface temperature at that location. Although the total amount of energy per year that both planets receive is the same, its different temporal and spatial distribution leads to extremely different climatic conditions.

\subsection{Sensitivity to asymmetric mean irradiation and heat capacity}

In all numerical experiments with eccentric orbits we assumed that the total radiation received over one year is symmetrically distributed among the two hemispheres ($\omega = 0^{\circ}$). Our results show, however that the expansion of the HZ due to seasonal variability is primarily determined by the peak of the locally incoming radiation and whether this maximal radiation suffices to lead to temporally ice-free regions. We therefore expect that an uneven distribution of mean radiation and a higher peak radiation at one pole, both associated with $\omega \neq 0^{\circ}$, leads to a further extension of the HZ. In order to quantify this effect, we consider the most extreme case $\omega = 90^{\circ}$ (solstice aligned with periastron). In a previous study \citep{Dressing_2010} experiments with $\omega = 30^{\circ}$ showed the maximal extent of the HZ. We also include one simulation with $\omega = 30^{\circ}$, although we expect that the value of $\omega$ associated with the maximal expansion of the HZ is sensitive to further parameters such as the planet's heat capacity.

\begin{figure}[htb!]
\centering \includegraphics[width=0.49\textwidth]{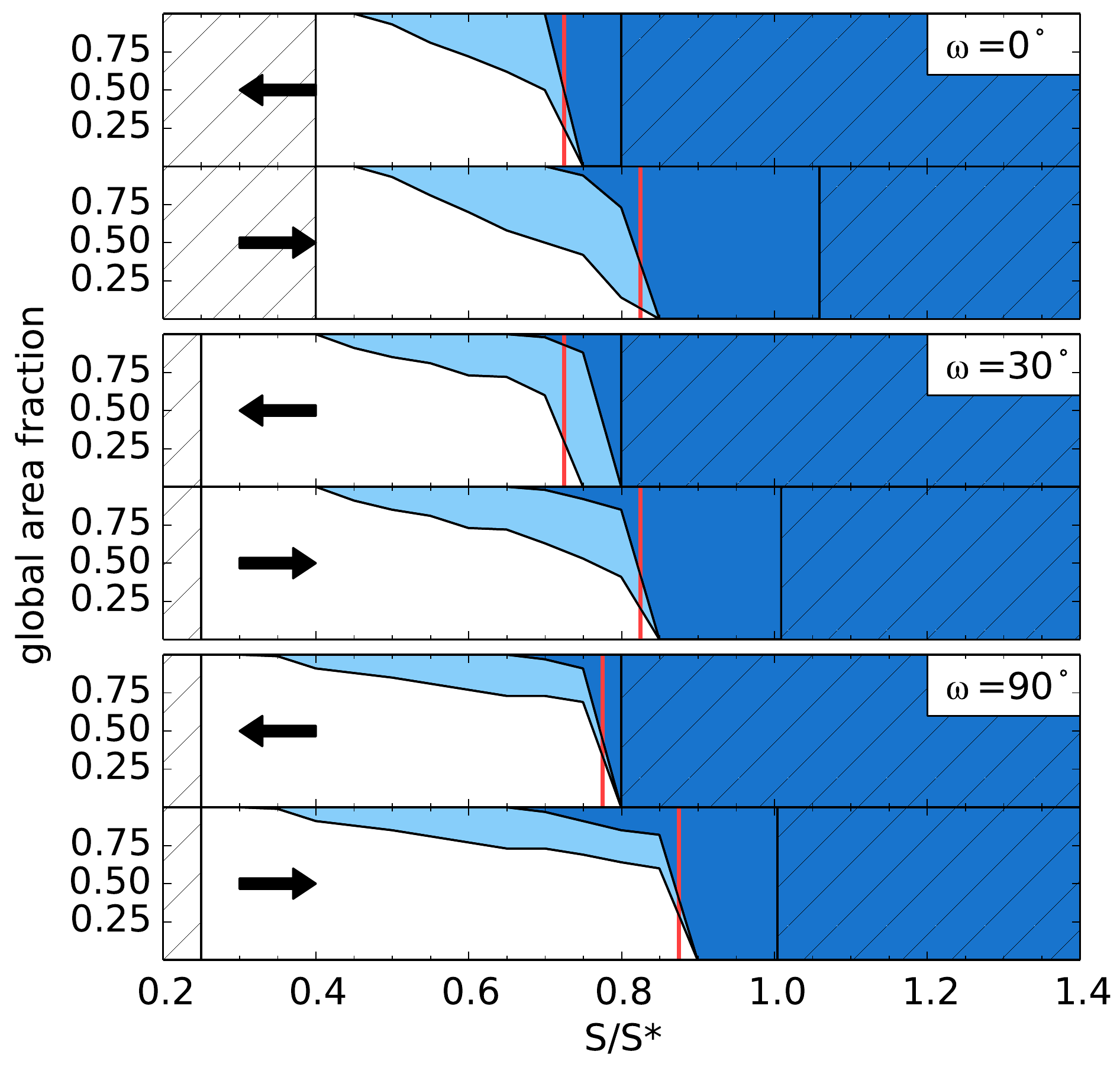}
\caption{Degree of habitability as in Figure \ref{fig:hab_degree}, here for $e = 0.5$, $\theta = \SI{90}{\degree}$, and $\omega = 0, 30, \SI{90}{\degree}$; the arrows indicate the direction of change of $\tilde{S}$, i.e. whether an ice-free or ice-covered initial state is assumed; the red vertical line marks the point of transition between the warm and cold state; hashes indicate areas outside the simulated range of $\tilde{S}$}
\label{fig:hab_degree_mvelp}
\end{figure}

The results are shown in Figure \ref{fig:hab_degree_mvelp}. As a consequence of the very warm polar summer for $\omega = \SI{90}{\degree}$, high latitudes becomes ice-free out to $\tilde{S} > 0.35$. Moreover, as the incoming radiation is less evenly distributed, the temporally ice-free areas shrink compared to $\omega = 0^{\circ}$. In contrast to \citet{Dressing_2010}, our experiments show the maximal expansion of the HZ for $\omega = \SI{90}{\degree}$ (Figure \ref{fig:hab_degree_mvelp}). We conclude that an uneven irradiation with $\omega = \SI{90}{\degree}$ leads to the most extreme climates in our experiments with small areas becoming temporary ice-free for even very low values of $\tilde{S}$.

\begin{figure*}[htb!]\centering
\includegraphics[width=0.49\textwidth]{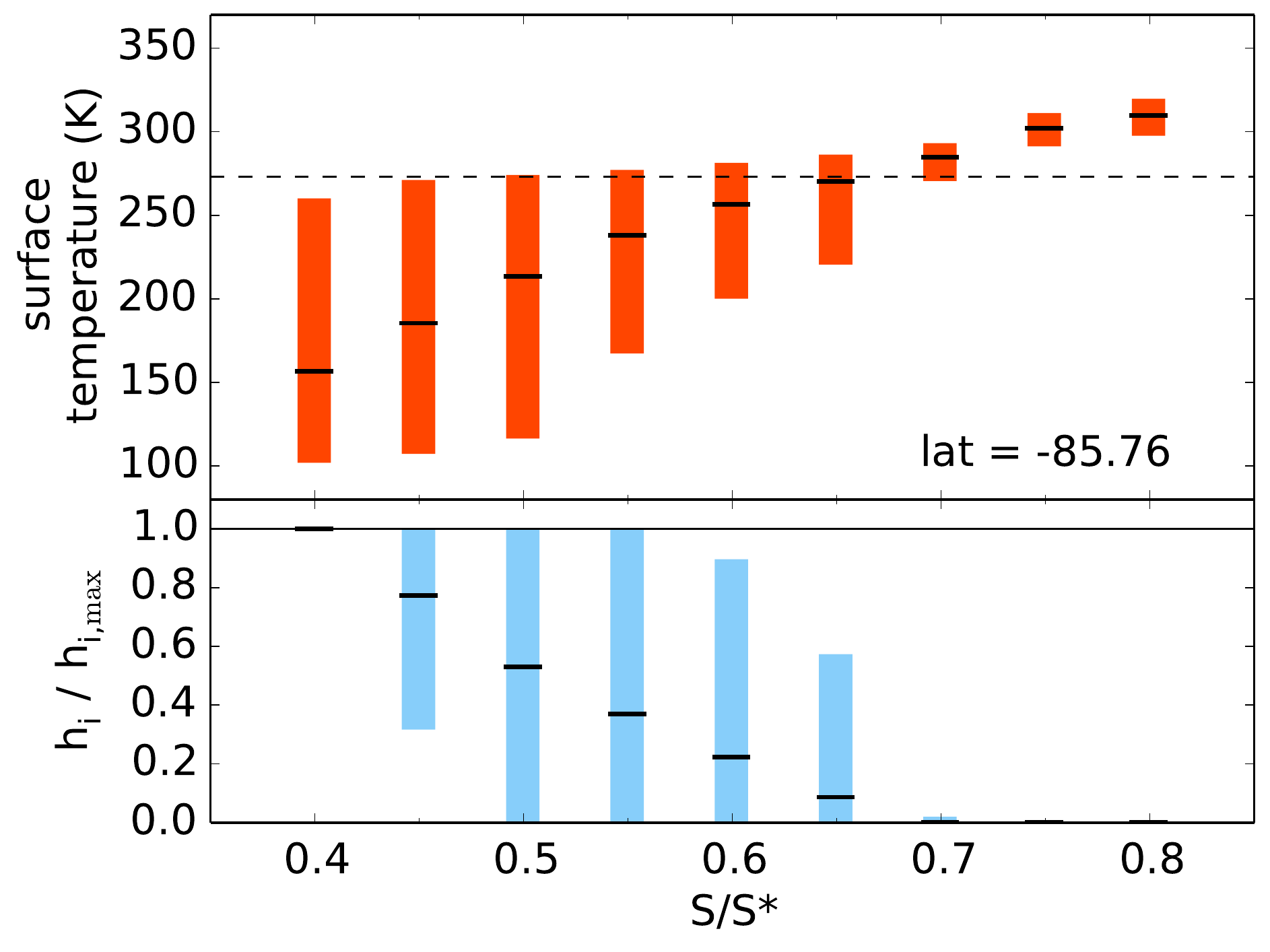}
\includegraphics[width=0.49\textwidth]{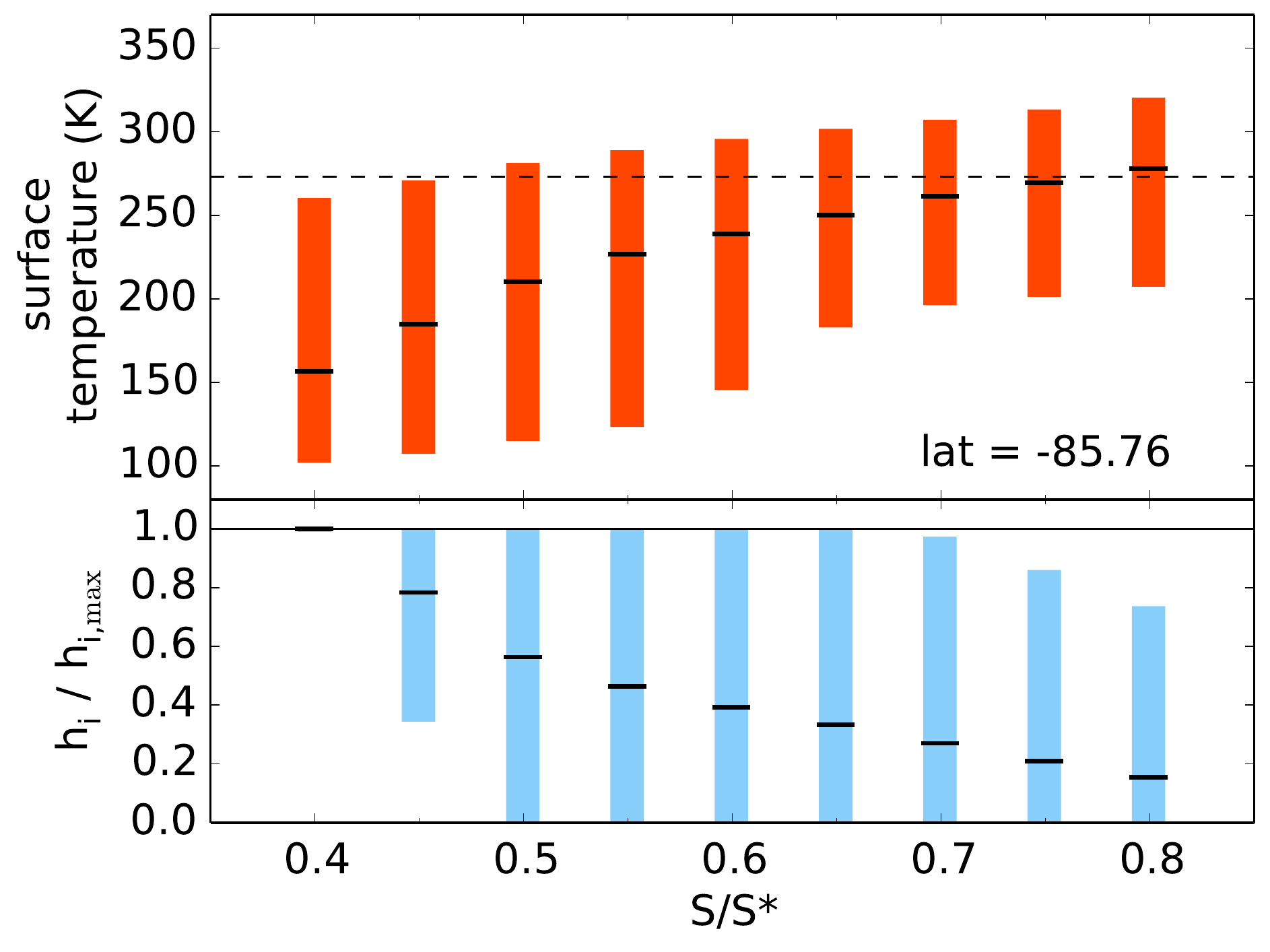}
\caption{Annual mean, maximum and minimum of surface temperature and normalized sea ice thickness ($h_{i,\rm{max}} = \SI{3}{m}$, indicated by the black horizontal line) of the experiments E05o90 with an ocean depth of 50 m (left) and E05o90oc10 (right) with an ocean depth set to 10 m (left)}
\label{fig:hab_sens_oc}
\end{figure*}

Variations of surface temperature due to seasonal forcing are damped by the heat capacity of the planet. The mean value and the amplitude of surface temperature at $e = 0.5$ and $\theta = 0^{\circ}$ and $\theta = 90^{\circ}$ are depicted in Figure \ref{fig:hab_sens_oc}. Both experiments show roughly the same pattern. In an ice-free state, temperature oscillations are relatively small ($\sim$ 10-20 K, Figure \ref{fig:hab_sens_oc}). With sea ice at the surface, this amplitude increases up to more than 100 K as the exchange of heat between the atmosphere and ocean is reduced and the ice-albedo feedback further amplifies temperature variations (Figure \ref{fig:hab_sens_oc}).

Choosing an ocean depth of 10 m in our sensitivity experiment hence affects the temperature oscillations and ice coverage only in warm climate states (Figure \ref{fig:hab_sens_oc}). In addition, lower heat capacity leads to the formation of ice at higher irradiation $\tilde{S}$, as less heat can be stored around periastron and the surface temperature thus drops below the freezing point around apoastron. Hence, although a reduced heat capacity reduces the continuously habitable surface area in favor of temporally habitable regions, it has no effect on the lower limit of $\tilde{S}$ that still leads to ice-free areas. The sensitivity experiment thus demonstrates that our estimate of the HZ is robust with respect to changing the heat capacity of the surface layer.

\section{Discussion}\label{sec:discuss}

In this work we provide for the first time a parametric investigation of the effect of seasonal variability on the climate of idealized Earth-like planets based on a general circulation model (GCM), having in mind the exploration of habitability conditions. Exploring the effects of the two parameters obliquity and eccentricity we build on previous work with energy balance models (EBMs) that focuses on the same parameters but in which simpler models are used \citep{Spiegel_2009,Dressing_2010}. At the same time our work complements parametric investigations with the same model focusing on different parameters \citep{Boschi_2013,Lucarini_2013}. Although the realism of our experiments regarding the determination of the HZ is constrained by both neglecting the silicate weathering cycle and the lack of a dynamic ocean and a dynamic sea ice model, we aim to provide valuable insights into the effects of seasonal variability for the climate of idealized planets on shorter time scales.

The numerical experiments performed in this work are based on the assumption of idealized aquaplanets. In order to test the robustness of our results further aspects need to be taken into account. First, our idealized planets are all characterized by uniform surface characteristics. As discussed in Section \ref{sec:model:planet}, surface heat capacity, albedo, and moisture supply can alter externally induced temperature oscillations and substantially affect the atmospheric dynamics of a planet. The effects of different land-sea distributions are taken into account in \citet{Williams_2003} and \citet{Spiegel_2009}, but a detailed investigation with general circulation models has not been done yet.

Second, our numerical experiments are performed with a slab ocean model and thus do not account for oceanic heat transport. On tidally locked planets, where strong spatial variations can imply spatially confined habitable conditions similar to our results \citep{Pierrehumbert_2011a}, oceanic heat transport can lead to a substantial extension of the HZ \citep{Hu_2013}. However,  for  high obliquity and low eccentricity \cite{Ferreira_2014} demonstrated that slab ocean models reproduce well the surface climate of coupled simulations for an ocean depth of 50 m.

Third,  only thermodynamic sea ice processes are represented in our model. Any dynamical processes such as wind induced sea ice drift or the flow of sea ice due to gravity are not taken into account. In some cases we expect this to limit the realism of our results. In the experiment of a planet on an eccentric orbit with $\theta = \SI{0}{\degree}$, for instance, the prescribed maximal sea ice thickness might bias our results towards temporally habitable conditions, because sea ice is not allowed to grow beyond a thickness of $\SI{3}{m}$ anywhere on the planet during apoastron. Without this restriction, sea ice could grow much thicker at the poles and subsequently flow towards lower latitudes, preventing the quick melt of the sea ice layer that appears in our simulation and thus resulting in a permanently unhabitable climate. Taking Earth as reference, a sea ice thickness beyond 3 m might indeed be realistic on exoplanets, as it is estimated that during Snowball events Earth was covered by an ice layer several hundreds of meters thick  \citep{Abbot_2010,Pierrehumbert_2011b}. A further limitation of our sea ice model might be due to a too high amplitude of the seasonal cycle of surface temperatures, attributable to the low vertical resolution of the ice layer in our model \citep{Abbot_2010}. Because associated high peak values of surface temperature are required for the existence of ice-free areas in some of our simulations, also this limitation of our sea ice model tends to bias our results towards (temporally) habitable conditions.

Fourth, the effect of a different rotation rate needs to be investigated because of its large impact on the meridional heat transport \citep{Vallis_2009}. Experiments with energy balance models where the effect of the planetary spin rate is parametrized through the meridional heat transport diffusion coefficient show that on planets with seasonal variability, the effect on habitability is comparatively small \citep{Spiegel_2009,Dressing_2010}. Besides its effect on the meridional transport of energy, the planetary spin also affects climate through diurnal variability, which may have a large effect on habitability for slowly rotating and phase-locked planets.

\begin{figure}[htb!]
\centering \includegraphics[width=0.5\textwidth]{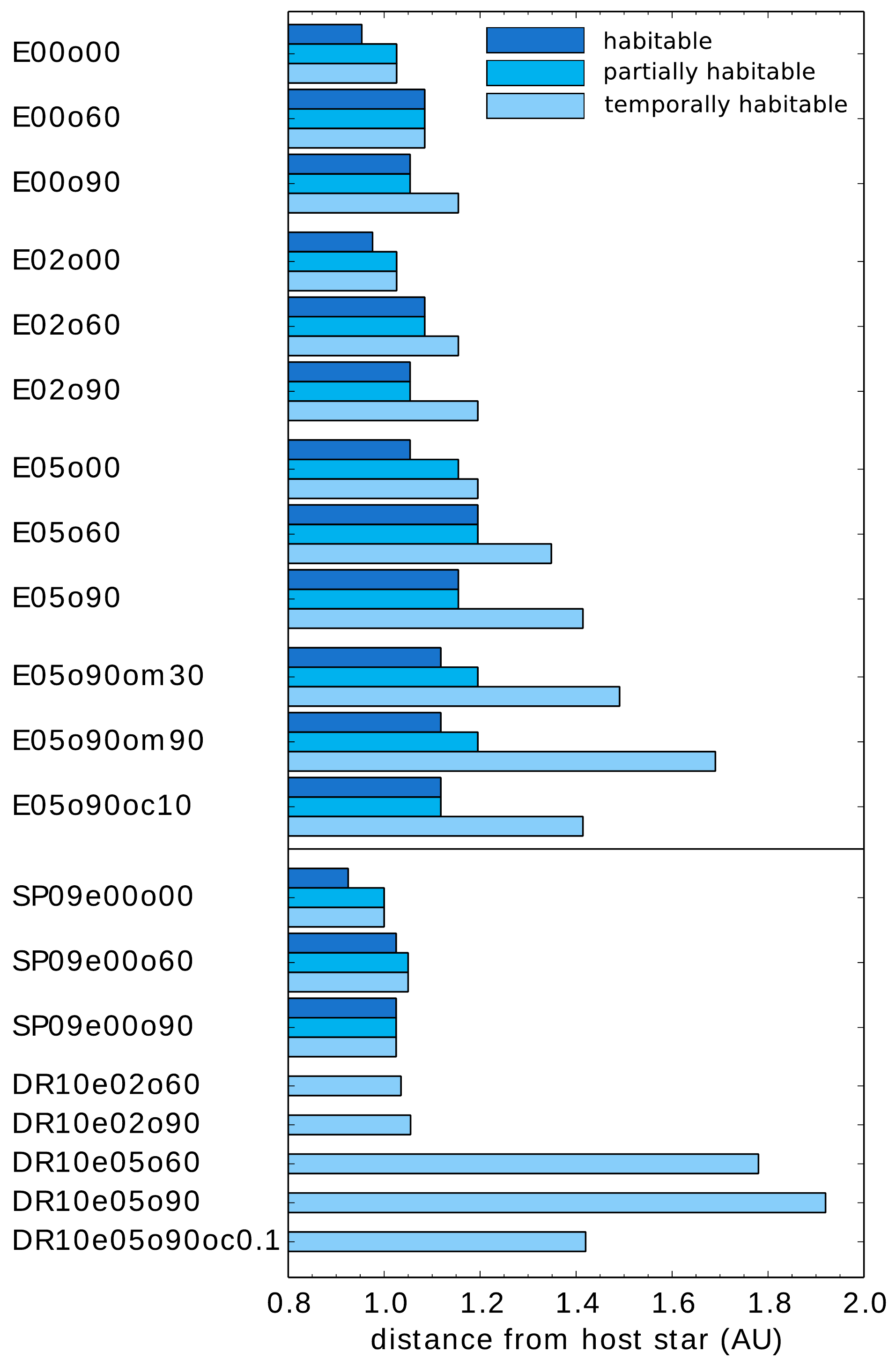}
\caption{Maximal distance between planet and host star that allows for habitable, partially habitable and temporally habitable climate conditions; the results of the last eight experiments denoted by SP09 and DR10 are taken from \citet{Spiegel_2009} and \citet{Dressing_2010} respectively; both use the same EBM and simulate the climate of Earth-like planets on circular (SP09) and eccentric orbits (DR10); while in both studies an ocean surface fraction of 70 \% is default, this fraction is set to 10 \% in the simulation DR10e05o90oc0.1}
\label{fig:zone}
\end{figure}

Several authors have addressed the effect of seasonal variability on the habitability of exoplanets using climate models of lower complexity. On circular orbits, our results are in good agreement with previous results based on one-dimensional latitudinal resolved EBMs (\citet{Spiegel_2009}, Figure \ref{fig:zone}). They also agree well in a qualitative sense with recent work based on an energy balance model \citep{Armstrong_2014}, showing a positive relation between obliquity and the maximal habitable zone. On eccentric orbits with $e=0.5$, however, simulations with EBMs tend to overestimate the outer boundary of the HZ relative to our results (\citet{Dressing_2010}, Figure \ref{fig:zone}). Despite this overestimation, the qualitative effect of obliquity agrees well also on eccentric orbits, showing the maximal extent for planets with $\theta = 90^{\circ}$ (Figure \ref{fig:zone}).

Conversely to the overestimation of the HZ by EBMs at $e=0.5$, simulations at $e=0.2$ show a smaller extent of the HZ simulated by EBMs than observed in our experiments (Figure \ref{fig:zone}). This somewhat surprising observation might be attributable to slightly different experimental setups of the EBMs used in the experiments at $e=0$ and $e=0.2$ \citep{Spiegel_2009,Dressing_2010}, but can also hint at a non-linear effect of eccentricity on the extent of the HZ in the EBMs. Such effect might originate from the complex interplay between the different timescales of variability, which is dramatically simplified by assessing only its consequence for the extent of the HZ.

Apart from apparent differences between the two different classes of models, such as the explicit simulation of the atmospheric circulation in our model, further explanations for the discrepancies are different definitions of habitability and the representation of ice processes. While \citet{Spiegel_2008} and subsequent studies use $T_{s} \geq \SI{273}{K}$ as condition for habitability, our definition is based on ice-coverage of the surface. Hence it suffices that the surface temperature shortly exceeds the melting point to be habitable in their model, while in our experiments it is required that the sea ice be locally completely melted. This difference becomes more severe due to the representation of the ice-albedo feedback in the two models. While the sea ice model of PlaSim considers only the areal extent of sea ice, i.e. a complete melting of the ice column is required to affect the surface albedo and therefore amplify the initial signal, the EBM incorporates a smooth dependency of surface albedo on temperature, which can be interpreted as the inclusion of melting ponds on top of the ice sheet and facilitates numerical computations.

The most critical limitations of our work are shared with these previous studies. These limitations are due to the neglection of the effect of CO$_2$ accumulation on the extent of the HZ. The most common estimates of the habitable zone rely on zero dimensional radiation convection models (RCMs). As opposed to our general circulation model, these models include feedback mechanisms of the carbon cycle: on the one hand, a decrease of surface temperature reduces the rate of chemical weathering at the surface, which leads to an accumulation of CO$_{2}$ in the atmosphere and an enhanced greenhouse effect. On the other hand, a higher concentration of CO$_{2}$ leads to an increased Rayleigh scattering of incoming radiation, which tends to further decrease the surface temperature. The outer boundary of the HZ, also called the maximum greenhouse limit, is then defined by the balance between these two processes \citep{Kasting_1993}. Because our aim is a parametric investigation of the effects of obliquity and eccentricity on habitability, we have to limit the number of parameters and our experiments can not provide a realistic estimate of habitability of a generic planet. In leaving aside the stabilizing effect of silicate weathering, however, we provide conservative estimates of the HZ.

\section{Conclusions} \label{sec:concl}

Our results show a great diversity of climates due to different obliquities and orbit eccentricities. Moreover, seasonal variability has a large effect on habitability. For planets with Earth-like atmospheres, seasonal variability can extend the maximal distance between planet and host star that still allows for habitable conditions from \SI{1.03}{AU} (obliquity $\theta = 0^{\circ}$, eccentricity $e = 0$) to \SI{1.69}{AU} ($\theta = 90^{\circ}$, $e = 0.5$, $\omega = 90^{\circ}$). While the effect of obliquity on habitability is comparatively small on circular orbits, it becomes highly relevant on eccentric orbits. This effect of seasonal variability on habitability is primarily due to regions that are ice-free only at some time of the year. An appropriate assessment of the HZ therefore asks for a clear distinction between different degrees of habitability (Table \ref{tab:hab}).

\begin{table*}[htb]\centering \footnotesize
\begin{tabular}{ l  r  r  r } \toprule & \multicolumn{3}{l}{Maximal distance from host star}\\ \cmidrule{2-4} Experiment & Habitable & Partially habitable & Temporally habitable \\ \midrule E00o00       & 0.95 AU & 1.03 AU & 1.03 AU \\ E00o60       & 1.08 AU & 1.08 AU & 1.08 AU \\ E00o90       & 1.05 AU & 1.05 AU & 1.15 AU \\ E02o00       & 0.98 AU & 1.03 AU & 1.03 AU \\ E02o60       & 1.08 AU & 1.08 AU & 1.15 AU \\ E02o90       & 1.05 AU & 1.05 AU & 1.20 AU \\ E05o00       & 1.05 AU & 1.15 AU & 1.20 AU \\ E05o60       & 1.20 AU & 1.20 AU & 1.35 AU \\ E05o90       & 1.15 AU & 1.15 AU & 1.41 AU \\ E05o90om30   & 1.12 AU & 1.20 AU & 1.49 AU \\ E05o90om90   & 1.12 AU & 1.20 AU & 1.69 AU \\ E05o90oc10   & 1.12 AU & 1.12 AU & 1.41 AU \\ \bottomrule \end{tabular}
\caption{Maximal distance between planet and host star that allows for habitable, partially habitable, and temporally habitable climatic conditions}
\label{tab:hab}
\end{table*}

Moreover, our experiments show that the multistability of the climate state is strongly influenced by the obliquity of a planet, in particular on circular orbits. The largest extent of the warm state is found for an obliquity around $\theta = \SI{55}{\degree}$, where the temperature distribution is most homogeneous. The smallest extent of the warm state and the largest extent of the cold state is found for planets with low values of obliquity. On eccentric orbits, the range of distances that allow for two stable climate states shrinks relative to circular orbits, possibly leading to monostability for planets with very large seasonal variability. While the extent of the warm climate state is a good proxy for the habitability of planets without seasonal variability, this does not hold for planets with high obliquities and on eccentric orbits, since also cold states can allow for ice-free regions. Whether life can exist in such extreme climates with strong temporal and spatial variability in surface conditions is still an open question \citep{Kane_2012}. Our results also suggest a need for more sophisticated indicators of climate that account for the effects of strong variability.

Some of the effects of seasonal variability can be attributed to an increase of the annual mean irradiation with eccentricity. Nonetheless, while this effect of eccentricity can in part explain the extension of the warm state on eccentric orbits, it does neither capture the widening effect on the HZ nor the limiting effect on the extent of the cold state.

Estimates of the outer boundary of the HZ that are based on RCMs and account for silicate weathering yield a maximal planet-star distance of \SI{1.67}{AU} \citep{Kasting_1993}, recently updated to \SI{1.70}{AU} \citep{Kopparapu_2013}. These estimates, however, rely on the assumption of a constant and uniform surface temperature, which is - as show in our results - a crude simplification of the problem. Planets with high seasonal variability can indeed feature a mean surface temperature below 240 K although having large areas completely ice-free throughout the year (Figure 8). How weathering might be affected by such a high spatial and temporal variability and what this implies for estimates of the outer boundary of the habitable zone is an interesting open question. Future investigations with GCMs that combine seasonal variability, a dynamic CO$_{2}$ concentration, and ocean and sea ice dynamics are required in order to provide further insights into the implications of seasonal variability for the climate of exoplanets and their ability to host life.

\section*{Acknowledgments}\label{sec:ackn}

The authors thank two anonymous reviewers for their  insightful comments on the first draft of this manuscript.  This research has received funding from the project NAMASTE - Thermodynamics of the Climate System, supported by the European Research Council under the European Community's Seventh Framework Programme (FP7/2007- 2013)/ERC Grant agreement No. 257106 and by the cluster of excellence Integrated Climate System Analysis and Prediction (CliSAP).

\bibliography{main}

\end{document}